\documentclass[preprint,12pt,number,square,sort&compress]{elsarticle}

\usepackage{graphicx}
\usepackage[export]{adjustbox}
\graphicspath{ {Figures/} }
\usepackage{fnpct}
\usepackage[colorlinks]{hyperref}
\usepackage{amsmath}
\usepackage{amsxtra}
\usepackage{amstext}
\usepackage{amssymb}
\usepackage{latexsym}
\usepackage{dsfont}
\usepackage{textcomp}
\usepackage{ulem}
\usepackage{multirow}
\usepackage{color}
\usepackage{epstopdf}
\usepackage{caption, subcaption}
\usepackage{bm}
\usepackage{soul}
\usepackage{natbib}
\usepackage{hyperref}
\hypersetup{
	colorlinks=true,
	linkcolor=cyan,
	filecolor=magenta,      
	urlcolor=blue,
}

\journal{Wave Motion}

\begin{document}

\newcommand{\edt}[1]{\textcolor{black}{#1}}  
\newcommand{\pd}[2]{\partial_{#1}#2}
\newcommand{\pdx}[1][]{\pd{x}#1}
\newcommand{\pdt}[1][]{\pd{t}#1}
\newcommand{\eps}{\varepsilon}
\newcommand{\sig}{\sigma}
\newcommand{\kap}{\kappa}
\newcommand{\gam}{\gamma}
\newcommand{\om}{\omega}
\newcommand{\bbet}{\bm{\beta}}
\newcommand{\bpsi}{\bm{\psi}}
\newcommand{\bkap}{\bm{\kap}}
\newcommand{\zp}{z^{+}}
\newcommand{\zm}{z^{-}}
\newcommand{\kp}{k^{+}}
\newcommand{\km}{k^{-}}
\newcommand{\kpj}{\kp_{j}}
\newcommand{\kmj}{\km_{j}}
\newcommand{\cp}{c^{+}}
\newcommand{\cm}{c^{-}}
\newcommand{\Vol}{\mathcal{V}}
\newcommand{\Sur}{\mathcal{S}}
\newcommand{\TM}{\mathsf{T}}
\newcommand{\SM}{\mathsf{S}}
\newcommand{\Oh}{\mathcal{O}}
\newcommand{\beginsupplement}{%
	\setcounter{table}{0}
	\setcounter{figure}{0}
	\setcounter{section}{0}
	\setcounter{subsection}{0}
	\setcounter{equation}{0}
	\setcounter{paragraph}{0}
	\setcounter{page}{1}
	\renewcommand{\thetable}{S\arabic{table}}%
	\renewcommand{\thefigure}{S\arabic{figure}}%
	\renewcommand{\thesection}{S\arabic{section}}%
	\renewcommand{\thesubsection}{S\arabic{section}.\arabic{subsection}}%
	\renewcommand{\theequation}{S\arabic{equation}}%
	\renewcommand{\theparagraph}{S\arabic{paragraph}}%
	\renewcommand{\thepage}{S\arabic{page}}%
	\renewcommand{\theHtable}{S\the\value{table}}%
	\renewcommand{\theHfigure}{S\the\value{figure}}%
	\renewcommand{\theHsection}{S\the\value{section}}%
	\renewcommand{\theHsubsection}{S\the\value{subsection}}%
	\renewcommand{\theHequation}{S\the\value{equation}}%
	\renewcommand{\theHparagraph}{S\the\value{paragraph}}%
}

\begin{frontmatter}
	
\title{Overall constitutive description of symmetric layered media based on scattering of oblique SH waves}
\author{Alireza V. Amirkhizi and Vahidreza Alizadeh}
\address{Department of Mechanical Engineering\\
University of Massachusetts, Lowell\\
One University Avenue, Lowell, MA 01854\\
\href{mailto:alireza\_amirkhizi@uml.edu}{alireza\_amirkhizi@uml.edu}}
\date{\today}

\begin{abstract}
This papers investigates the scattering of oblique shear horizontal (SH) waves off finite periodic media made of elastic and viscoelastic layers. \edt{It further considers whether a Willis-type constitutive matrix (in temporal and spatial Fourier domain) may reproduce the scattering matrix (SM) of such a system. In answering this question the procedure to determine the relevant overall constitutive parameters for such a medium is presented.} To do this, first the general form of the dispersion relation and impedances for oblique SH propagation in such coupled Willis-type media are developed. The band structure and scattering of layered media are calculated using the transfer matrix (TM) method. The dispersion relation may be derived based on the eigen-solutions of an infinite periodic domain. \edt{The wave impedances associated with the exterior surfaces of a finite thickness slab are extracted from the scattering of such a system.} Based on reciprocity and available symmetries of the structure and each constituent layer, the general form of the dispersion and impedances may be simplified. The overall quantities may be extracted by equating the scattering data from TM with those expected from a Willis-type medium. It becomes evident that \edt{a Willis-type coupled constitutive tensor with components that are assumed independent of wave vector} is unable to reproduce all oblique scattering data. Therefore, non-unique \edt{wave vector dependent} formulations are introduced, whose SM matches that of the layered media exactly. It is further shown that \edt{the dependence of the overall constitutive tensors of such systems on the wave vector} is not removable even at very small frequencies and incidence angles and that analytical considerations significantly limit the potential forms of the spatially dispersive constitutive tensors.
\end{abstract}


\end{frontmatter}

\section{Introduction}

Spatial dispersion, i.e. the wave-vector dependence of wave propagation, has been widely studied in electromagnetism, crystal optics, and photonic crystals, but has received relatively little attention in stress wave propagation in solids \cite{Pitaevskii1984,agranovich_crystal_1984}. In electromagnetism, such phenomena are rooted in the possible dependence of the electric displacement field at a given point, $\mathbf{D}$, on the electric field, $\mathbf{E}$, in a neighborhood around that point, i.e. nonlocal physics. Another mathematical approach to such phenomena is to consider the dependence of $\mathbf{D}$ on spatial derivatives of $\mathbf{E}$, which in the case of wave propagation, or after performing a spatial Fourier transform, can be written in terms of $\mathbf{k}$ functionality. \edt{The general non-local behavior is  not always representable in a wave-vector based formalism particularly for finite systems. However, the focus of this paper is to determine whether the overall scattering response of certain heterogeneous systems may be fully represented using overall constitutive tensors in Fourier domain, and to describe the process of extracting the relevant components of such tensors.} Similarly, for stress waves a limited representation of non-locality after spatial Fourier transform is written as the elastic constants exhibiting a dependence on the wave vector besides the frequency, i.e. $C_{ijkl}(\omega,\mathbf{k})$. The 4th order tensorial nature of the elastic constants introduces more extensive mathematical complexity in stress wave problems. Many interesting features in EM of conductive materials and metamaterials, such as virtual surface waves \cite{Maradudin1973}, metasurfaces \cite{Asadchy,Diaz-Rubio}, cloaking \cite{Shalin2015}, longitudinal polarization waves \cite{Melnyk1968,JONES1969,Melnyk1970,Ruppin1975}, negative refractions \cite{Agranovich2006,Chern2013}, anti-resonance response \cite{AndreaAi2011,Alitalo2013} among others, are either due or related to spatial dispersion. Extensive studies has been also done on different types of metamaterials in which spatial dispersion is a necessity to characterize the system due to its periodicity (particularly when wavelengths of interest are comparable with unit cell dimensions) \cite{Li,Vehmas2014,Papantonis,Belov2003,Rockstuhl2008,Tyshetskiy2014,Miret2016,Yves,Ushkov}. Experimentally, Hopfiled et al. \cite{Hopfield1963} observed spatial dispersion behavior in optical properties of crystals and showed that classical optics of nonmagnetic crystal theory could not accurately predict their empirically obtained scattering data. In another study by Portigal et al. \cite{Portigal}, it was shown that by considering the first-order spatial dispersion ``acoustical activity" can occur in crystals which exhibit optical activity. Higher order theories have also been utilized in order to exactly satisfy the impedance matching for all incident angles in perfectly matched layers (PML), \cite{Tretyakov1998} or to homogenize for overall properties of metamaterials with nonmagnetic inclusions \cite{Ciattoni2015}. Agarwal et al. \cite{Agarwal} proposed a theory related to the nature of electromagnetic field in spatially dispersive media. They concluded that without the introduction of any additional ad-hoc boundary conditions, the scattering problem of a plane-parallel slab can be solved completely within the framework of classical electromagnetic theory. This theory can be treated as a step forward in understanding and formulating of the spatial dispersion phenomenon in general. Puri et al. \cite{Puri1983} also demonstrated that the optical properties of finite bounded crystals, e.g. GaAs, CuCl, etc, show nonlocal behavior for frequencies near an exciton resonance. In another study on bianisotropic media Belov et al. \cite{Belov} proved that for a 2D lattice of infinitely long parallel conducting helices, strong spatial dispersion is observable even at very low frequencies along the spiral axis of the bianisotropic medium. Yaghjian et al. \cite{Yaghjian2013} developed an anisotropic homogenization theory for spatially dispersive periodic arrays based on Maxwell's equations. They mathematically proved that it is impossible to characterize metamaterials formed by periodic arrays of polarizable inclusions without spatially dispersive representation of permittivity and inverse transverse permeability.

In contrast, nonlocal stress wave propagation in elastic composites has not received much attention in the literature.  A number of authors have looked at the extraction of the overall mechanical properties from scattering data. Popa et al. \cite{Popa2009} showed that the anisotropic effective mass density tensor components, which were extracted from the scattering data, can be independently controlled by properly designing the inclusions geometry in a host fluid. To enhance the applicability of retrieval methods, Castanie et al. \cite{Castanie2014} analyzed the problem of extracting the overall properties of an orthotropic slab, whose normal is not one of the material principal axes. Park et al. \cite{Park2016} studied the behavior of anisotropic acoustic metamaterials by considering non-diagonal effective mass density tensor in the presence of spatial dispersion. Lafarge and Nemati \cite{Lafarge2013,Nemati2014} developed a nonlocal theory for sound propagation in fluid-filled rigid frame porous media. The nonlocal approach was further expanded to include the effect of arrays of Helmoholtz resonators in \cite{Nemati2015} \edt{and general non-local treatment of dissipative phononic liquids in} \cite{nemati_nonlocal_2017}. In these works, the focus has been on propagation of pressure acoustic waves and the effect of microstructure on effective bulk modulus and density. Lee et al. \cite{Lee2016}, extended the work in \cite{Park2016} to mechanical metamaterials, where the elasticity tensor components, beyond just the bulk modulus, are incorporated. Recently a mathematically exact theory for Willis-type media has been developed along with a matching field integration technique for normal incidence in \cite{Amirkhizi2017}, where transfer matrix method (TMM) is used to calculate the scattering exactly. The field averaging technique in \cite{Amirkhizi2017} is based on the explicit integration of the wave equations and it matches the scattering response exactly as well. The method is applied to viscoelastic and non-symmetric structures without any modification. It must be noted that Nemat-Nasser \cite{Nemat-Nasser2015a} developed a variational approach to extract the band structure of SH waves in periodic composite media, where group velocity and energy flux vectors were calculated explicitly. In another study by Srivastava \cite{Srivastava2016}, SH waves at an interface between a homogeneous material and a layered periodic composite were investigated. The laminate had a periodically layered structure in $x_1$-direction, with interface between the homogeneous medium and layered structure was normal to the $x_2$-axis and the SH waves were polarized along the $x_3$-axis. Negative refraction and beam steering was observed in a wide range of frequency. 

In this work, the scattering approach of \cite{Amirkhizi2017} is used for the oblique incidence of SH waves in layered media to calculate the scattering matrix (SM) without any numerical (e.g. FE) effort. \edt{It is determined whether a Willis-type constitutive tensor formalism in spatial Fourier domain can reproduce the scattering response of such systems, and if so, how to retrieve these constitutive parameters.} Transfer matrix methods has been widely and successfully used to calculate exact and approximate scattering for oblique SH waves; See, for example, Vinh et al. \cite{Vinh2015}. To simplify the derivations, the work is focused on symmetric unit cells and structures. In the following sections, first, the general form of the dispersion relations and impedances are extracted by solving the equations of motion, taking into account the Willis-type constitutive law. The results are discussed in light of various symmetry and reciprocity requirements. A general approach for extracting the overall constitutive parameters from the scattering data (dispersion and impedances) is developed. Next, a two-phase symmetric layup composite is presented to apply the general approach. It is shown that the response may not be fully represented \edt{by constitutive matrices that are independent of wave vector}, even when one allows for coupling terms in the Willis formalism. \edt{When the constitutive tensors are allowed to be functions of wave vector,} the SH wave propagation results are not enough to fully determine all constitutive functions. Two special representations are discussed: one with a diagonal constitutive tensor, and \edt{one with the diagonal terms that are not functions of the wave vector (only off-diagonal components are allowed to change with the wave vector)}. A third representation is studied for which \edt{it is shown that analytical considerations eliminate any possible compatible solution, pointing out that further limitations on the potential acceptable forms of the constitutive tensors exist.} Based on this, it is expected that by considering all other propagation modes (P and SV) some of the potential solutions will further be ruled out. This approach and the treatment of asymmetric structures will be taken up in future efforts. 

\section{SH wave dispersion, slowness, and impedance in coupled media}
The elastodynamics of source-free media may be derived based on the combination of the \edt{compatibility} and equilibrium equations
\begin{gather} 
		(\pd{i}{v_j} + \pd{j}{v_i})/2 = \pdt{\eps_{ij}}, \label{eq:cont} \\
		\pd{j}{\sig_{ji}} = \pdt{p_{i}}, \label{eq:equil}
\end{gather}
\edt{with the constitutive law.} Here $v, \sig, p$, and $\eps$ denote particle velocity, stress, momentum density (per unit volume), and strain, respectively, and subscripts after $\partial$ represent partial differentiation with respect to that variable. For any material or composite that has translational symmetry along an axis, e.g. $x_3$, SH-waves with particle velocity polarization along this symmetry axis and propagation direction in the $x_1x_2$ plane normal to it may be studied independently if all constituents have sufficient symmetry to decouple such shear deformation from other tensor components of the dynamic and kinematic quantities in Equations~\eqref{eq:cont} and \eqref{eq:equil}, which can then be collected as
\begin{equation}
	\begin{pmatrix}
	0 & \pd{2} & \pd{1} \\
	\pd{2} & 0 & 0 \\
	\pd{1} & 0 & 0
	\end{pmatrix}
	\begin{pmatrix}
	v_3 \\
	\tau_4 \\
	\tau_5
	\end{pmatrix}
	=
	\pdt \begin{pmatrix}
	p_3 \\
	\gam_4 \\
	\gam_5
	\end{pmatrix}, \label{eq:wave}
\end{equation}
where the Voigt notation is used, i.e. $\gam_4 = 2 \eps_{23} = 2 \eps_{32}, \gam_5 = 2 \eps_{31} = 2 \eps_{13}, \tau_4 = \sig_{23} = \sig_{32}, \tau_5 = \sig_{31} = \sig_{13}$. \edt{In the following only systems are considered, for which it is assumed that a Willi-type formalism in temporal and spatial Fourier domain is capable of reproducing their overall (scattering) behavior. This could be a homogeneous, layered, finite, or infinite system, and the validity of such assumption is established by applying this formalism and inspecting its actual success in this reproduction.} To study oblique incidence of SH waves, only a portion of the full Willis-type constitutive law is used
\begin{equation}\label{eq:const}
	\begin{pmatrix}
	v_3 \\
	\tau_4 \\
	\tau_5
	\end{pmatrix}
	=
	\begin{pmatrix}
		\eta_{33} & \kap_{34} & \kap_{35}\\ 
		\kap_{43} & \mu_{44} & \mu_{45} \\
		\kap_{53} & \mu_{54} & \mu_{55}
	\end{pmatrix}
	\begin{pmatrix}
		p_3\\ 
		\gam_4 \\
		\gam_5
	\end{pmatrix}, 
\end{equation}
where $\eta_{33}$ represents the appropriate specific volume, $\mu_{ij}, i, j = 4, 5$ are the relevant moduli of elasticity (e.g. $\mu_{12} = C_{2331}$), while $\kap_{ij}$ denote particle velocity/strain and stress/momentum density couplings. \edt{All components are functions of the frequency and wave vector in general.} Equation~\eqref{eq:const} is a slightly transformed version of the formulation used in  \cite{Milton2007, Willis2009}  in that stress is grouped with particle velocity instead of momentum density \cite{Amirkhizi2017}. Matrix inversion of Equation~\eqref{eq:const} gives the form in terms of density and compliance. One reason for using this grouping is the similarity with Equations~\eqref{eq:cont} and \eqref{eq:equil}, \edt{in that momentum density is grouped with strain components in a single column vector, and particle velocity is grouped with stress components.} Furthermore and as discussed in \cite{Amirkhizi2017} the integration of wave equations also leads to volume integral definitions for overall strain and momentum density quantities, while overall stress and particle velocities have natural definitions as surface integrals.
In frequency domain, the physical quantities described above ($\beta = v_3, \tau_j, p_3, \gam_j, j = 4, 5$) are written as $\beta(x,t) = \Re(\beta_{c}(x) e^{-i\om t})$, where $\om = 2 \pi f$ is the angular frequency and $\beta_{c}$ is the complex amplitude. For a spatial Fourier component $\beta_{c}(x) = \beta_{c0} e^{i \bm{k} \cdot \bm{x}}$\footnote{The subscripts $_c$ and $_{c0}$ will be dropped in the following unless there is potential for confusion.}, where $\bm{k}$ is the wave vector, equation~\eqref{eq:wave} can be written as 
\begin{equation}
	-\frac{1}{\om}\begin{pmatrix}
		0 & k_2 & k_1 \\
		k_2 & 0 & 0 \\
		k_1 & 0 & 0
	\end{pmatrix}
	\begin{pmatrix}
		v_3 \\
		\tau_4 \\
		\tau_5
	\end{pmatrix}
	=
	\begin{pmatrix}
		p_3 \\
		\gam_4 \\
		\gam_5
	\end{pmatrix}.
\end{equation}
Using the column vector $\bbet = (v_3, \tau_4, \tau_5)^T$, these equations can be combined as
\begin{gather}
	-  \bkap \bpsi \bbet = \bbet.
\end{gather}
Here $\bkap$ is the Willis-type constitutive matrix in Equation~\eqref{eq:const} and 
\begin{equation}
	\bpsi = 
	\begin{pmatrix}
	0 & s_2 & s_1 \\
	s_2 & 0 & 0 \\
	s_1 & 0 & 0
	\end{pmatrix} =
	\frac{1}{\om} 
	\begin{pmatrix}
	0 & k_2 & k_1 \\
	k_2 & 0 & 0 \\
	k_1 & 0 & 0
	\end{pmatrix},
\end{equation}
is made of slowness vector components $\bm{s} = \bm{k}/\om$. The dispersion relation between the wave vector components, $k_j$, $j = 1, 2$, and the frequency $\om$ can now be found from
\begin{equation}
	\det(\bm{I} +  \bkap \bpsi) = 0.
\end{equation}
In the most general form, the dispersion surface equation will look like
\begin{equation}
\begin{split}
	0 = D(\om, k_1, k_2) = \om^2 &+ \om((\kap_{35}+\kap_{53})k_1 + (\kap_{34}+\kap_{43})k_2) \\
	&+ (\kap_{34}\kap_{53}+\kap_{43}\kap_{35}-\eta_{33}(\mu_{45}+\mu_{54}))k_1 k_2 \\
	&+(\kap_{35} \kap_{53} - \eta_{33}\mu_{55}) k_1^2 +(\kap_{34} \kap_{43} - \eta_{33}\mu_{44}) k_2^2. \label{eq:disp}
\end{split}
\end{equation}
For any 2 selected variables, e.g. wave vector components $k_1$ and $k_2$, or even one variable and a ratio, e.g. frequency $\om$ and slowness component $s_2$, there may exist multiple points on the complex dispersion surface, which will be formally identified by a superscript, e.g. $\om^\alpha = \om^\alpha (k_1, k_2)$ or $s_1^\alpha = s_1^\alpha(\om, s_2)$. While this allows for finding the general solutions, in most cases of interest one or more of the variables are restricted. For example, steady-state cases are limited to real frequencies. \edt{For example, in particular cases of interest in this paper, i.e. scattering of oblique SH waves in systems that have translational invariance along the $x_2$ axis, a single value of $k_2$ is fixed everywhere in the system. Furthermore, expectation of analyticity of constitutive tensors (except at potentially discrete locations) may be utilized to conclude that for most of the wave vector space, there are limited number of solutions to \eqref{eq:disp}, in fact only 2. This is in agreement with the physical observation of overall scattering of layered media, calculated further below. Moreover, any potential point for which the analyticity assumption fails should be treated carefully and separately. Note that the Bloch multiplicity of $k_1$ in these cases does not provide a physical distinct new solution; this is further discussed below.} For each solution $\alpha$ \edt{(in the following indexed by $\pm$ for 2 solutions)}, the null space of the matrix $\bm{I} + \bkap \bpsi$ is generated by vectors $\bbet^\alpha = (1, -z_{43}^\alpha, -z_{53}^\alpha)^T$, where $z_{j3}^\alpha$, $j = 4, 5$, are formally defined as impedances associated with the points $\alpha$ on the dispersion surface. 

\paragraph{Reciprocity and symmetries}

A reversal in time coordinate, $t \rightarrow -t$ (along with $\om \rightarrow -\om$) would not change the physics of reciprocal systems and in particular the shape of their dispersion surfaces. However, such a transformation will change the solutions of the dispersion relation~\eqref{eq:disp} unless the coefficient of $\om$ vanishes, therefor: \footnote{In a system where constitutive tensors are function of wave vector, this will lead to a single yet more complex condition, instead of the two independent ones in the text.}
\begin{align}
	\kap_{34} + \kap_{43} = 0, \label{eq:k43} \\
	\kap_{35} + \kap_{53} = 0. \label{eq:k53}
\end{align}
\edt{Note that the analyticity considerations of the previous paragraph is implicitly used here, as this discussion is based on considering the lowest order approximation of constant constitutive parameters in a small enough neighborhood around the point of interest on the dispersion surface (which is also assumed to be one with the normal topology observed almost everywhere as shown later in the example results).} The dispersion relation may now be slightly simplified, e.g. in slowness space
\begin{equation}
	(\eta_{33} \mu_{55} + \kap_{35}^2) s_1^2 + (\eta_{33} \mu_{44} + \kap_{34}^2) s_2^2 + (\eta_{33}(\mu_{45} + \mu_{54}) + 2 \kap_{34} \kap_{35}) s_1 s_2 = 1.
\end{equation}
If the system has a mirror or $\pi$ rotational symmetry in or around the $x_1$ or $x_2$ directions then the coefficient of $k_1 k_2$ should vanish as well 
\begin{equation}
	\eta_{33}(\mu_{45} + \mu_{54}) + 2 \kap_{34} \kap_{35} = 0, \label{eq:k1k2}
\end{equation}
and the slowness/dispersion formulas will simplify further to 
\begin{equation} \label{eq:disp_s}
(\eta_{33} \mu_{55} + \kap_{35}^2) s_1^2 + (\eta_{33} \mu_{44} + \kap_{34}^2) s_2^2 = 1.
\end{equation}
Even without the mirror symmetry simplification, for a reciprocal medium, the impedances have the relatively compact forms
\begin{align}
	z_{43} &= \mu_{44} s_2 + \mu_{45} s_1 + (\kap_{34}/\eta_{33}) (1 + \kap_{34} s_2 + \kap_{35} s_1), \label{eq:zg43} \\
	z_{53} &= \mu_{55} s_1 + \mu_{54} s_2 + (\kap_{35}/\eta_{33}) (1 + \kap_{35} s_1 + \kap_{34} s_2). \label{eq:zg53}
\end{align}
Mirror and rotational symmetries of the material or structure will further restrict the constitutive tensors, particularly when they are considered as functions of the wave vector.\footnote{As pure SH waves are discussed here, at least mirror symmetry in the $x_3$ direction is needed. Note that none of the quantities considered here are pseudo-tensors. Therefore, mirror symmetries on planes normal the other two axes are essentially equivalent to $\pi$ rotations around them. If there is no mirror symmetry normal to the $x_3$ axis, it is unlikely to get a pure SH wave without couple stresses and spins.} For example if $x_2 \rightarrow -x_2$ is a symmetry operation of the system, then $z_{43}$ should switch sign under $k_2 \rightarrow -k_2$, while $z_{53}$ should stay unchanged. A sufficient\footnote{And necessary except at discrete locations on the dispersion surface, e.g. $s_1 = 0$.} condition set for this is
\begin{equation}
	\mu_{45} = \mu_{54} = \kap_{34} = 0. \label{eq:x2sym}
\end{equation}
Similarly if $x_1 \rightarrow -x_1$ is a symmetry operation then
\begin{equation}
	\mu_{45} = \mu_{54} = \kap_{35} = 0. \label{eq:x1sym}
\end{equation}
However, for systems where the coupled constitutive parameters are functions of the wave vector, the conditions will take the form of parity restraints on the functional $k_1$ or $k_2$ dependence of the coupling constants. If $x_2 \rightarrow -x_2$ is a symmetry operation, then $\mu_{44}, \mu_{55}, \eta_{33}, \kap_{35}$ are even functions of $k_2$, while $\mu_{45}, \mu_{54}, \kap_{34}$ are odd. If $x_1 \rightarrow -x_1$ is a symmetry operation, then $\mu_{44}, \mu_{55}, \eta_{33}, \kap_{34}$ are even functions of $k_1$, while $\mu_{45}, \mu_{54}, \kap_{35}$ are odd. 

\section{Transfer matrix of layered structures for oblique SH waves}
The transfer matrix of a material layer normal to the $x_1$-axis, and identified by index $l$, for SH waves with velocity polarization along $x_3$ axis relates the particle velocity and traction components on the two boundaries:
\begin{equation}
	\begin{pmatrix}
		v_3(x_{1,l+1})\\ 
		\tau_5(x_{1,l+1})
	\end{pmatrix} 
	= 
	\TM_{l}
	\begin{pmatrix}
		v_3(x_{1,l})\\ 
		\tau_5(x_{1,l})
	\end{pmatrix}.
\end{equation}
$\TM_{l}$ is a function of the frequency and/or wave vector components, the layer's constitutive parameters, and its thickness $d_{l} = x_{1,l+1} - x_{1,l}$, where $x_{1,l}$ and $x_{1,l+1}$ represent the coordinates of the two surfaces of the layer. To satisfy continuity along all interfaces at all times, the frequency and the $k_2$ component of the wave vector should match across all layers. For each such pair, the homogeneous wave equation has the general solution superposing two independent SH waves
\begin{equation}
	\begin{pmatrix}
		v_3(x_1, k_{1,l}^\alpha) \\
		\tau_5 (x_1, k_{1,l}^\alpha)
	\end{pmatrix}
	=
	\begin{pmatrix}
		1 \\
		-z_{53,l}^\alpha(k_{1,l}^\alpha)
	\end{pmatrix}
	A_{l}^\alpha(k_{1,l}^\alpha) e^{i k_{1,l}^\alpha x_1}, 
	\quad 
	x_{1,l} \leq x_1 \leq x_{1,l+1}
\end{equation}
where $\alpha = +, -$ represents the two solutions, $A_{l}^\alpha(k_{1,l}^\alpha)$ are the complex amplitudes of the two waves, the common phase $e^{i (k_2 x_2 -\om t)}$ is dropped, and 
\begin{equation} \label{eq:z53}
	z_{53,l}^\alpha(k_{1,l}^\alpha) = -\frac{\tau_5 (x_1, k_{1,l}^\alpha)}{v_3(x_1, k_{1,l}^\alpha)}, 
\end{equation}
is the impedance of layer $l$ associated with the wave vector $k_{1,l}^\alpha$.
For a symmetric and reciprocal system, the transfer matrix can be written as:
\begin{equation} \label{eq:tmsym}
	\TM = 
	\begin{pmatrix}
		\cos k_1 d & -\dfrac{i}{z_{53}} \sin k_1 d \\
		-i z_{53} \sin k_1 d & \cos k_1 d
	\end{pmatrix}.
\end{equation}

Using the continuity of the particle velocity and traction vectors, one may calculate the transfer matrix of a cell consisting of $n_l$ layers as 
\begin{equation} \label{eq:tmcell}
	\TM_{c} = \TM_{n_l} ... \TM_{2} \TM_{1}, 
\end{equation}
and that of a finite slab consisting of $n_c$ cells as
\begin{equation}
	\TM_{s} = (\TM_{c})^{n_c}.
\end{equation}
Consider an elastic unit cell that generates an infinitely periodic structure. The band structure of such a system may calculated by solving the eigenvalue problem
\begin{equation} \label{eq:eign}
	\TM_{c} \bm{\zeta} =
	 e^{ik_{1}^\alpha d_{c}} \bm{\zeta},
\end{equation}
where $d_{c} = x_{1,n_l+1} - x_{1,1}$ and for finite impedances $\bm{\zeta} = (1, -z_{53}^\alpha(k_{1}^\alpha))^T$. Note the phase ambiguity in the overall wave vector component $k_{1}^\alpha$, where any integer multiple of $2\pi/d_c$ may be added to or subtracted from it. Enforcing continuity requirements on the wave vector allows one to remove this ambiguity \cite{Amirkhizi2017}.

The scattering matrix of a multi-layered slab extending from $x_{1,a}$ to $x_{1,b} = x_{1,a} + d_s = x_{1,a} + n_c d_c$ and placed in between two half-spaces (also identified by subscripts $a$ and $b$) may be calculated from its transfer matrix and the properties of the two half spaces on by using the continuity conditions. For the sake of simplicity consider the two media to be identical and symmetric, with characteristic impedance $z_b = z_a = z_0$. When the sample is also symmetric and reciprocal and using the homogenized values in Equation~\eqref{eq:tmsym} \cite{Amirkhizi2017}:
\begin{equation} \label{eq:STM}
\begin{aligned}
\SM_{ba} = \SM_{ab} &= \frac{2}{2 \cos k_1 d_s  -i (z_{53}/z_0 + z_0/z_{53}) \sin k_1 d_s}, \\
\SM_{bb} = \SM_{aa} &= \frac{i (z_{53}/z_0 - z_0/z_{53}) \sin k_1 d_s}{2 \cos k_1 d_s - i (z_{53}/z_0 + z_0/z_{53}) \sin k_1 d_s}.
\end{aligned}	
\end{equation}
where $\SM_{ba}$ and $\SM_{aa}$ denotes transmission and reflection coefficients, respectively, in terms of kinematic quantities (particle velocity). The symmetry and reciprocity of the system removes the need for superscript $\alpha$ as for the only 2 SH solutions $z_{53}^+ = -z_{53}^-, k_1^+ = -k_1^-$. \edt{Note that point-wise inversion of the measured (or calculated) SM using these equations to determine $k_1$ and $z_{53}$ does not provide a unique solution for $k_1$. However, this ambiguity may be removed by enforcing the (physically motivated) continuity of $k_1$ as a function of frequency (similar to the traditional procedure of phase spectral analysis; see for example \cite{Nantasetphong2018}).} The inter-relation of $\SM$ and $\TM$ is presented in \cite{Amirkhizi2017}. Therefore, one may define the overall properties of a heterogeneous sample by finding the parameters $k_1$ and $z_{53}$ (assuming $d = d_s$, of course) with which Equation~\eqref{eq:tmsym} reproduces its exact transfer matrix. It can be shown that these definitions are independent of the number unit cells in the analysis \cite{Amirkhizi2017}.
If all layers have $x_2$ mirror symmetry (or $\pi$ rotational around $x_1$), the system does so as well, and therefore the simplifications discussed above may be used. That means in the approximation where the constitutive parameters are not functions of the wave vector, equations~\eqref{eq:x2sym} will hold. The dispersion relation and impedances will now simplify to:
\begin{gather}
	(\eta_{33} \mu_{55} + \kap_{35}^2) s_1^2 + \eta_{33} \mu_{44} s_2^2 = 1, \\
	z_{53} = \mu_{55} s_1 + (\kap_{35}/\eta_{33}) (1 + \kap_{35} s_1), \\
	z_{43} = \mu_{44} s_2.
\end{gather}
Furthermore, if the system has mirror symmetry normal to $x_1$ as well as every layer (or $\pi$ rotational around $x_2$ or $x_3$) then equations~\eqref{eq:x1sym} will also hold, enforcing a diagonal form for the constitutive tensor and  
\begin{gather}
\eta_{33} \mu_{55} s_1^2 + \eta_{33} \mu_{44} s_2^2 = 1, \\
z_{53} = \mu_{55} s_1, \\
z_{43} = \mu_{44} s_2.
\end{gather}

\paragraph{Reciprocity and symmetries with explicit dependence of constitutive tensor on the wave vector} When the constitutive tensor is assumed a function of the wave vector, the parity considerations should be used. The constitutive tensor does not  have to be diagonal, and in fact for asymmetric structures off-diagonal terms may be necessary. Note that $z_{43}$ is not accessible from scattering, and in fact $\tau_4$ and $v_3$ vary through the appropriate boundary in the unit cell with complicated profiles. With a reasonable definition or calculation of $z_{43}$ (e.g. considering scattering off an oblique cut or via integrating along the thickness direction), more equations will be available to determine constitutive parameters. Without further information however, at least one of the parameters may be considered free. In the following the simplifying Equations~\eqref{eq:k43}, \eqref{eq:k53}, and \eqref{eq:k1k2} are used. The normal component of slowness may be written as $s_1 = \pm s_1(\om, s_2)$ based on the dispersion relation and therefore any even function of $s_1$ can be written as an even function of $\om$ and any odd function of $s_1$ can be written as $sign(s_1)f(\om)$, with $f(\om)$ vanishing at zero frequency. With $x_1$ symmetry, and based on Equation~\eqref{eq:zg53}, $\kap_{35}$ will have this latter (odd) form, which may be chosen as a free parameter and set equal to zero. This is not feasible for a $x_1$-asymmetric structure, but it may be possible to get the value of $\kap_{35}$ for a normal incidence calculation and use throughout. The other quantities can now be fully determined based on the transfer matrix. With both $x_1$- and $x_2$-symmetries, and a first order approximation in $s_2$\footnote{$s_2$ has to be non-dimensionalized with some characteristic slowness $s_2/\bar{s}$ for a formal series consideration. But we neglect this process here since we only look at the first order approximation.}
\begin{equation}
\begin{split}
	\mu_{44} \approx \mu_{44,0}(\om), \quad \mu_{55} \approx \mu_{55,0}(\om), \quad \eta_{33} \approx \eta_{33,0}(\om), \\
	\mu_{54} (= -\mu_{45}) \approx s_2 \mu_{54,1}(\om), \quad \kap_{34} \approx s_2\kap_{34,1}(\om). 
\end{split} \label{eq:lin}
\end{equation} 
The left hand side of all thee approximations are general functions of $\om$ and $s_2$. Even though we have set $\kap_{35} = 0$, at each frequency there are infinitely many potential constitutive functions that could match the observed dispersion and impedances. Some cases in particular are of interest. First, one may assume that the constitutive matrix is diagonal. Even in this case, there is no need for $\eta_{33}$ to be a function of the wave vector and after enforcing a non-local form for it, one can calculate the diagonal functions as
\begin{gather}
	\mu_{55}(\om, s_2) = \frac{z_{53}(\om, s_2)}{s_1(\om, s_2)}, \label{eq:m55dd} \\
	\eta_{33}(\om) = \frac{1}{\mu_{55}(\om, 0)s_1(\om, 0)^2}, \label{eq:eta33sc1}\\
	\mu_{44}(\om, s_2) = \frac{1-\eta_{33}(\om)\mu_{55}(\om, s_2)s_1(\om, s_2)^2}{\eta_{33}(\om)s_2^2}. \label{eq:mu44sc1}
\end{gather}

The second potential choice is to enforce that all of the diagonal quantities are independent of the wave vector. In this case:
\begin{equation} \label{eq:z53_2}
	z_{53} = \mu_{55} s_1 + \mu_{54} s_2.
\end{equation}
$\mu_{55}$ and $\eta_{33}$ are easily determined as 
\begin{gather}
	\mu_{55}(\om) = \frac{z_{53}(\om, 0)}{s_1(\om, 0)}, \label{eq:mu55sc2} \\ 
	\eta_{33}(\om) = \frac{1}{\mu_{55}(\om)s_1(\om, 0)^2}. \label{eq:eta33sc2}
\end{gather}
Furthermore, 
\begin{equation} \label{eq:mu54}
	\mu_{54}(\om, s_2) = \frac{z_{53}(\om, s_2) - \mu_{55}(\om) s_1(\om, s_2)}{s_2}.
\end{equation}
To determine the other two functions, one may start with the dispersion relation
\begin{equation} \label{eq:disp_2}
	\eta_{33} \mu_{55} s_1^2 + (\eta_{33} \mu_{44} + \kap_{34}^2) s_2^2 = 1,
\end{equation}
then divide by $\eta_{33} \mu_{55}$, and implicitly differentiate twice to get
\begin{equation}
	s_1 s_1'' + s_1'^2 + \mu_{44}/\mu_{55} + [(\kap_{34}\kap_{34}'' + \kap_{34}'^2) s_2^2 + 4 \kap_{34} \kap_{34}' s_2 + \kap_{34}^2]/(\eta_{33} \mu_{55}) = 0, 
\end{equation}
where $'$ represents differentiation with respect to $s_2$. At $s_2 = 0$ this equation is simplified by noting that $\kap_{34}(\om, 0) = 0$, since it is an odd function of $s_2$, and $s_1'(\om, 0) = 0$, since $s_1(\om, s_2)$ is an even function of $s_2$. Therefore
\begin{equation}
	\mu_{44}(\om) = -\mu_{55}(\om) s_1(\om, 0) s_1''(\om, 0).
\end{equation} 
Finally
\begin{equation} \label{eq:kap^2}
	\kap_{34}^2(\om, s_2) = \frac{1 - \eta_{33}(\om) [\mu_{55}(\om)s_1(\om, s_2)^2 + \mu_{44}(\om) s_2^2]}{s_2^2}. 
\end{equation}
The particular root of $\kap_{34}$ may not be determined without further information (e.g. from $z_{43}$) but is irrelevant at this point. It is possible to evaluate the goodness of first order approximations $\mu_{54} (= -\mu_{45}) \approx s_2 \mu_{54,1}(\om)$ and $\kap_{34} \approx s_2\kap_{34,1}(\om)$ at this point. Similarly, one may evaluate the goodness of second order fits $\mu_{44} \approx \mu_{44,0} + s_2^2 \mu_{44,2}$ and $\mu_{55} \approx \mu_{55,0} + s_2^2 \mu_{55,2}$ in the diagonal description. In summary, in this scenario, $\kap_{34}^2$, $\eta_{33}$ and $\mu_{44}$ can only be obtained using dispersion relation, Equation~\eqref{eq:disp_2}, and $\mu_{55}$ and $\mu_{54}$ are merely extractable from impedance equation, Equation~\eqref{eq:z53_2}. This avoids any inconsistency that may occur if the equations are coupled. 

Such an inconsistency may occur if instead of assuming free parameter $\kap_{35} = 0$, one instead opts for selecting $\kap_{34}$ as the free parameter and set it to zero, while keeping the diagonal elements independent of the wave vector. In such a case, the dispersion relation in slowness domain, Equation~\eqref{eq:disp_s}, and impedance $z_{53}$, Equation~\eqref{eq:zg53}, will simplify further to:
\begin{gather}
	(\eta_{33} \mu_{55} + \kap_{35}^2) s_1^2 + \eta_{33} \mu_{44}s_2^2 = 1, \label{eq:disp_3} \\
	z_{53} = \mu_{55} s_1 + \mu_{54} s_2 + (\kappa_{35}/\eta_{33}) (1 + \kappa_{35} s_1). \label{eq:z53_3}
\end{gather}
Having in mind the $x_1$-symmetry of the system, which leads to $\kap_{35}$ being an odd function of $s_2$, the values of $\mu_{55}$ and $\eta_{33}$ can be determined exactly from Equations~\eqref{eq:mu55sc2}, \eqref{eq:eta33sc2}. However, the value of $\mu_{54}$ will be a function of $\kap_{35}$ as:
\begin{equation} 
	\begin{split}
		\mu_{54}(\omega,s_2) &= \{z_{53}(\omega,s_2) - \mu_{55}(\omega)s_1(\omega,s_2) \\
			&- [\kappa_{35}(\omega,s_2)/\eta_{33}(\omega)][1 + \kappa_{35}(\omega,s_2) s_1(\omega,s_2)]\}/s_2. \label{eq:mu54_sc3}
	\end{split}
\end{equation}
The equation for $\mu_{44}$ can be derived by differentiating the dispersion relation, Equation~\eqref{eq:disp_3}, twice with respect to $s_2$ and simplifying it at $s_2 = 0$:
\begin{equation}
	\mu_{44}(\omega) = -\mu_{55}(\omega)s_1(\omega,0)s_1''(\omega,0) - \frac{\kappa_{35}'^2(\omega,0) s_1^2(\omega,0)}{\eta_{33}(\omega)}.
\end{equation}
Both these quantities are written in terms of $\kap_{35}$ and its derivatives. In order to determine $\kap_{35}$, one can differentiate the dispersion relation one more time, i.e. a total of three times, to eliminate non-dispersive $\mu_{44}(\omega)$ and solve the resulting equation numerically. However, unlike the previous case where impedance $z_{53}$ and $\kap_{34}$ were uncoupled, $\kap_{35}$ appears in the impedance equation as well. A cumbersome analysis may be performed here leading to a necessary consistency condition between the dispersion and impedance equations, with the final form:
\begin{equation} \label{eq:self_cons}
	\begin{split}
		12z_{53}'^2(\omega,0)s_1(\omega,0)s_1''(\omega,0) &+ 4z_{53}'(\omega,0)z_{53}'''(\omega,0)s_1^2(\omega,0) +\\ [\mu_{55}(\omega)/\eta_{33}(\omega)]
		&\times[3{s_1''}^2(\omega,0) + s_1(\omega,0)s_1''''(\omega,0)]=0.
	\end{split}
\end{equation}
In the following a simple example is analyzed and for the sake of brevity here we state without reproducing the results that such a condition was violated almost everywhere in the frequency domain. Therefore, the potential free parametric choices stated earlier are indeed limited quite significantly by the physics of the problem. As a final case, note that if one considers either $\mu_{45} + \mu_{54} = 0$ as the free parameter and set it equal to zero, based on Equation~\eqref{eq:k1k2}, one would still require either of $\kap_{34}$ or $\kap_{35}$ to vanish as well, leading to one of the cases considered above. 

\section{Example: Determining the overall constitutive parameters}
\begin{figure}[!ht]
	\centering\includegraphics[height=220pt]{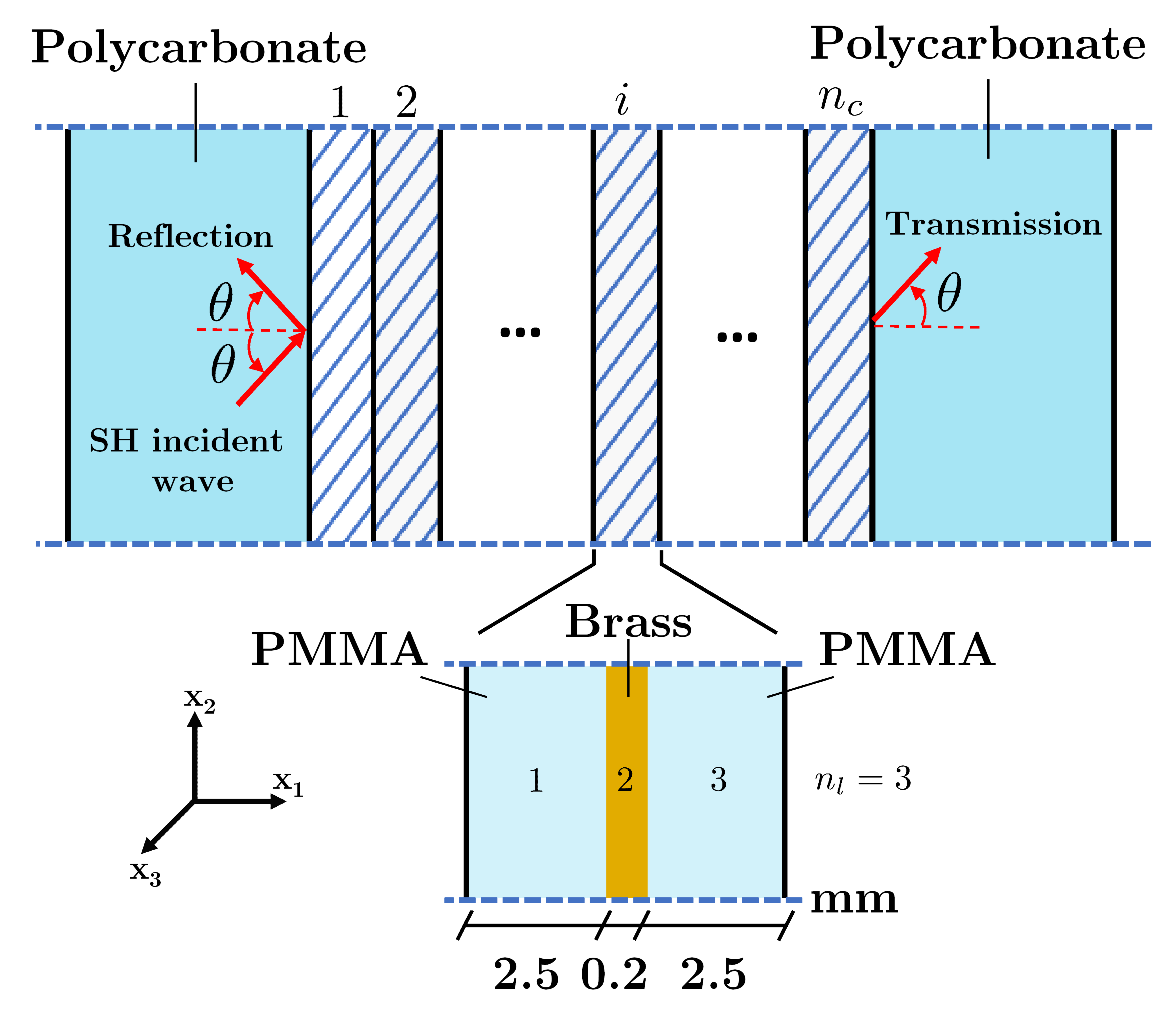}
	\caption{\label{fig:schcomp} The schematic view of the periodic layered system with a symmetric 2-phase unit cell of acrylic (PMMA) and brass. The incidence and transmission media are selected as polycarbonate (PC). Arrows show the oblique SH incident wave along with the reflection and transmission waves.}
\end{figure}

\begin{figure}[!ht]
	\begin{subfigure}[b]{0.5\linewidth}
		\centering\includegraphics[height=90pt,center]{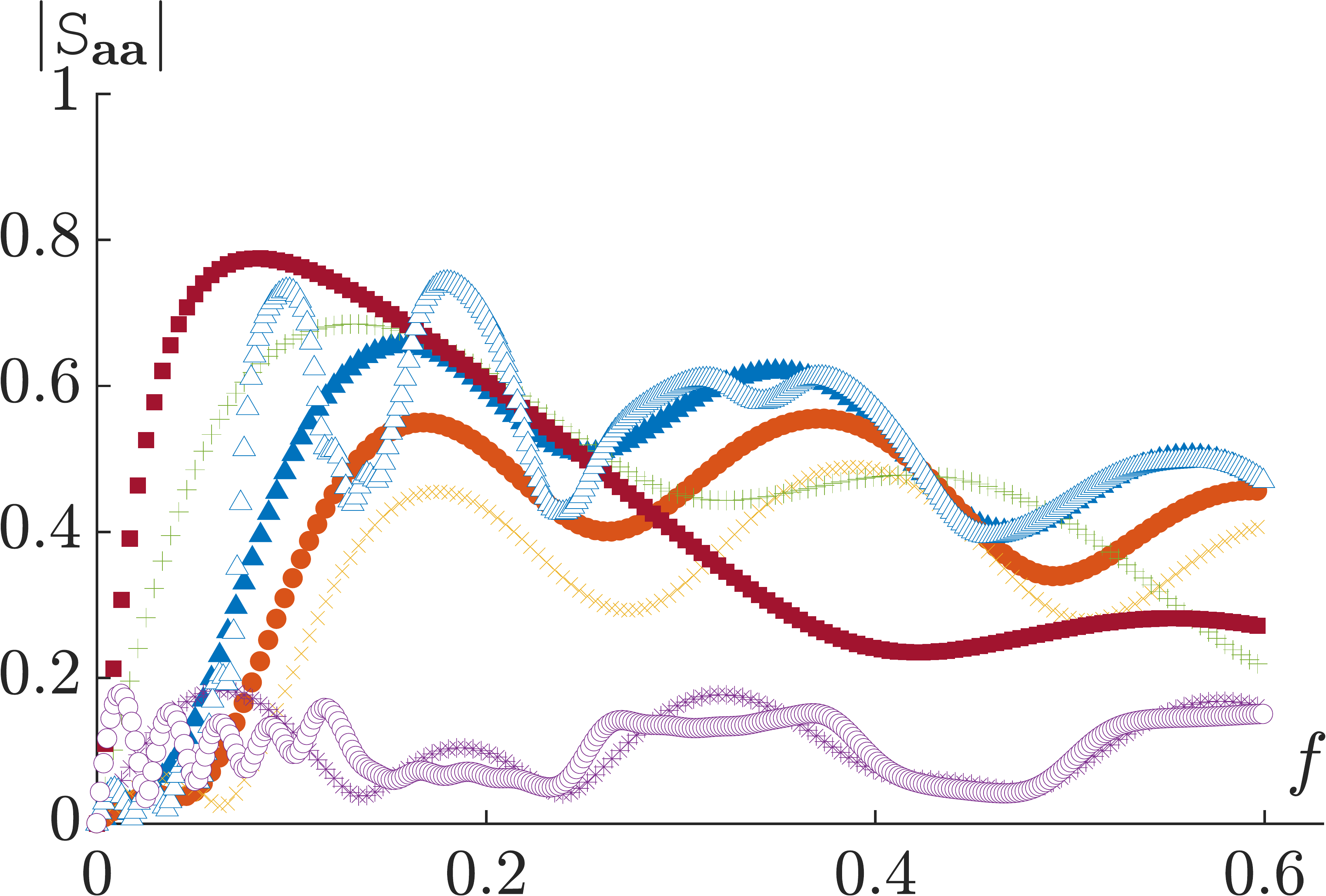}
		\caption{\label{fig:abs-Saa}}
	\end{subfigure}%
	\begin{subfigure}[b]{0.5\linewidth}
		\centering\includegraphics[height=90pt,width=195pt,left]{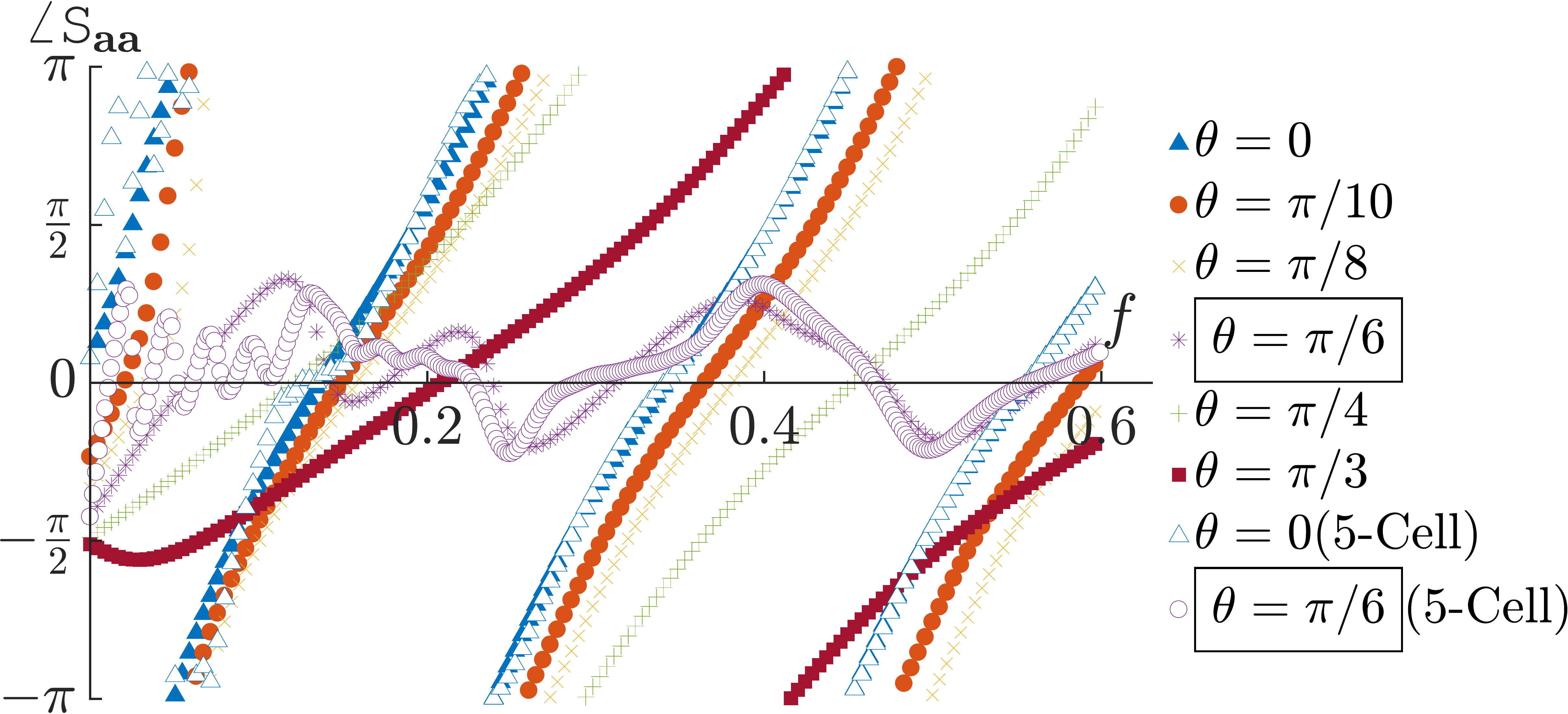}
		\caption{\label{fig:ang-Saa}}
	\end{subfigure}
	\begin{subfigure}[b]{0.5\linewidth}
		\centering\includegraphics[height=90pt,center]{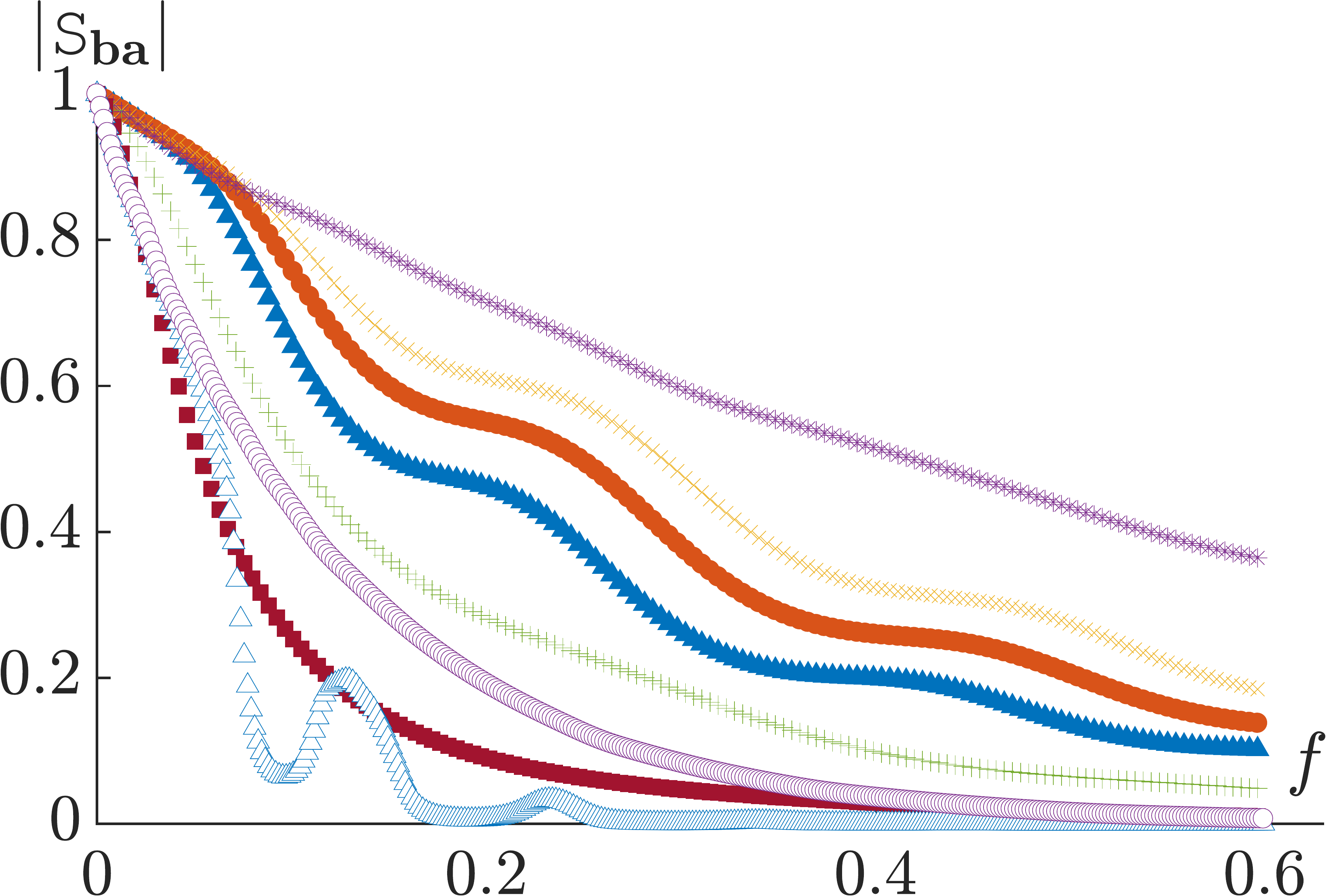}
		\caption{\label{fig:abs-Sba}}
	\end{subfigure}%
	\begin{subfigure}[b]{0.5\linewidth}
		\centering\includegraphics[height=90pt,width=195pt,left]{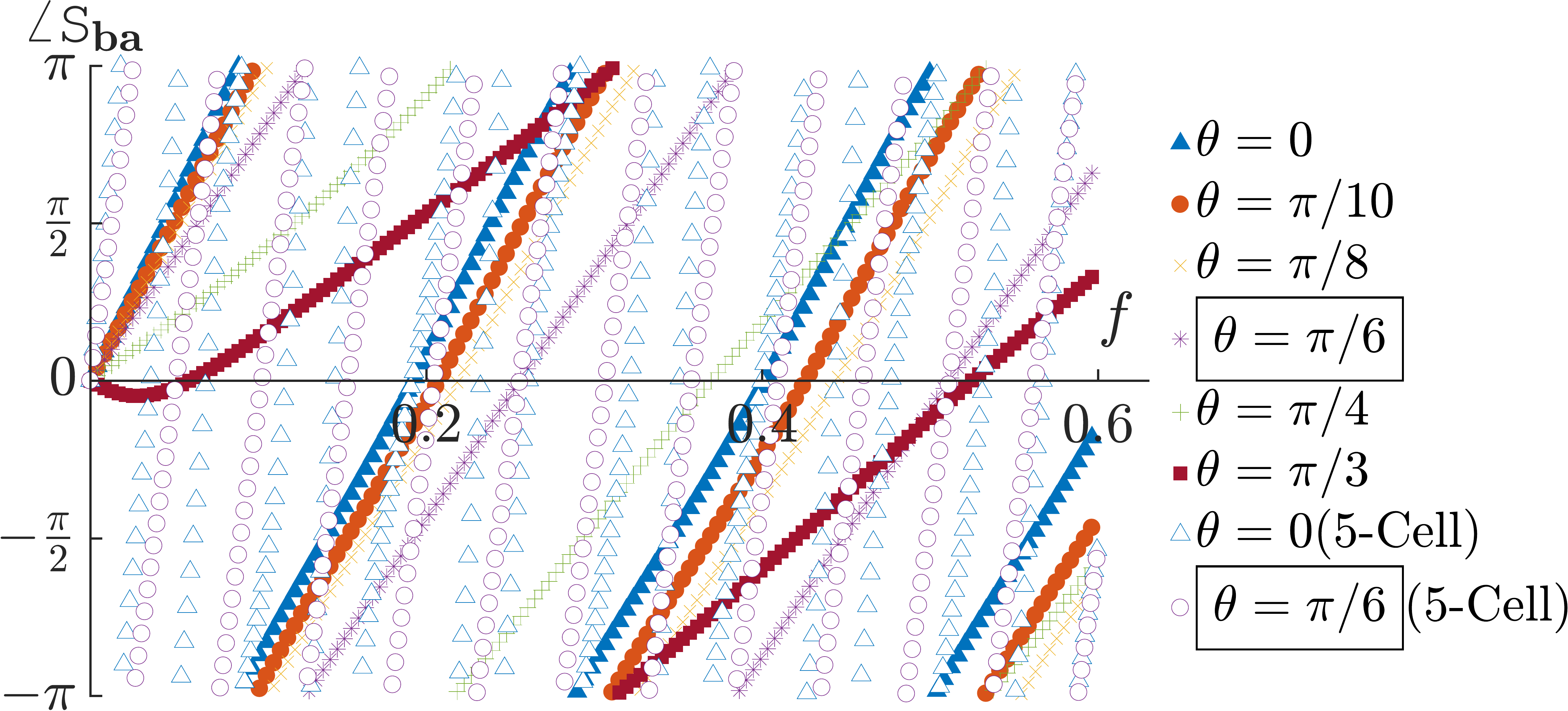}
		\caption{\label{fig:ang-Sba}}
	\end{subfigure}
	\caption{\label{fig:Scat}(\subref{fig:abs-Saa}) and (\subref{fig:abs-Sba}) show the magnitude of reflection and transmission, as functions of frequency, $f$, for different values of $s_2$ $(n_c=1)$. At $\theta=0, \pi/6$ graphs for 5 unit cells ($n_c=5$, hollow markers) are also shown. (\subref{fig:ang-Saa}) and (\subref{fig:ang-Sba}) show the phase angle of the same quantities. At ${\theta=\pi/6}$ incident angle, or $s_2 = 0.447$, the reflection is substantially lower in comparison to the other incident angles, giving the appearance of an impedance matched system.} 
\end{figure}

\begin{figure}[!ht]
	\centering
	\begin{subfigure}[b]{0.5\linewidth}
		\centering\includegraphics[height=95pt]{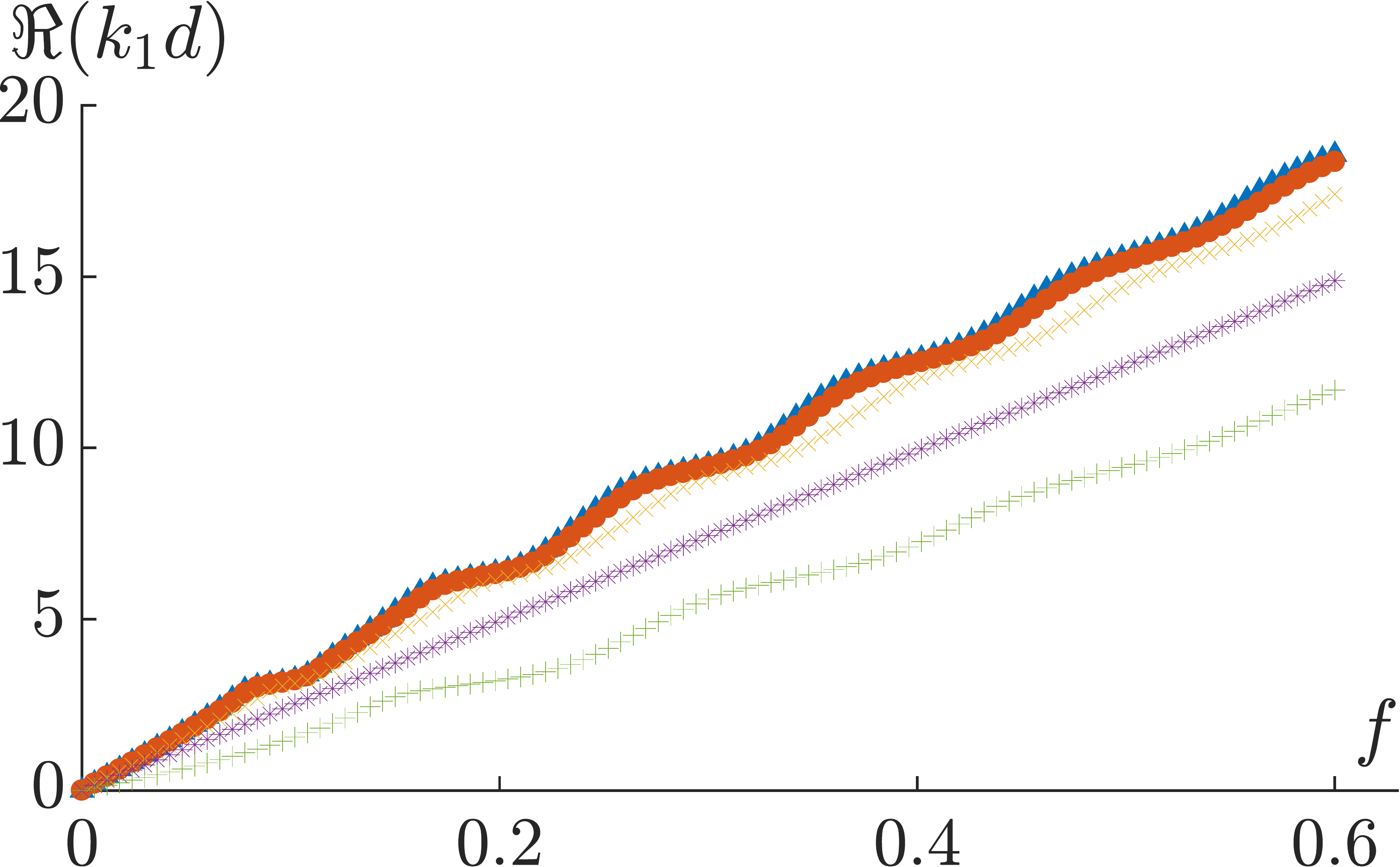}
		\caption{\label{fig:Re-k1d}}
	\end{subfigure}%
	\begin{subfigure}[b]{0.5\linewidth}
		\centering\includegraphics[height=95pt]{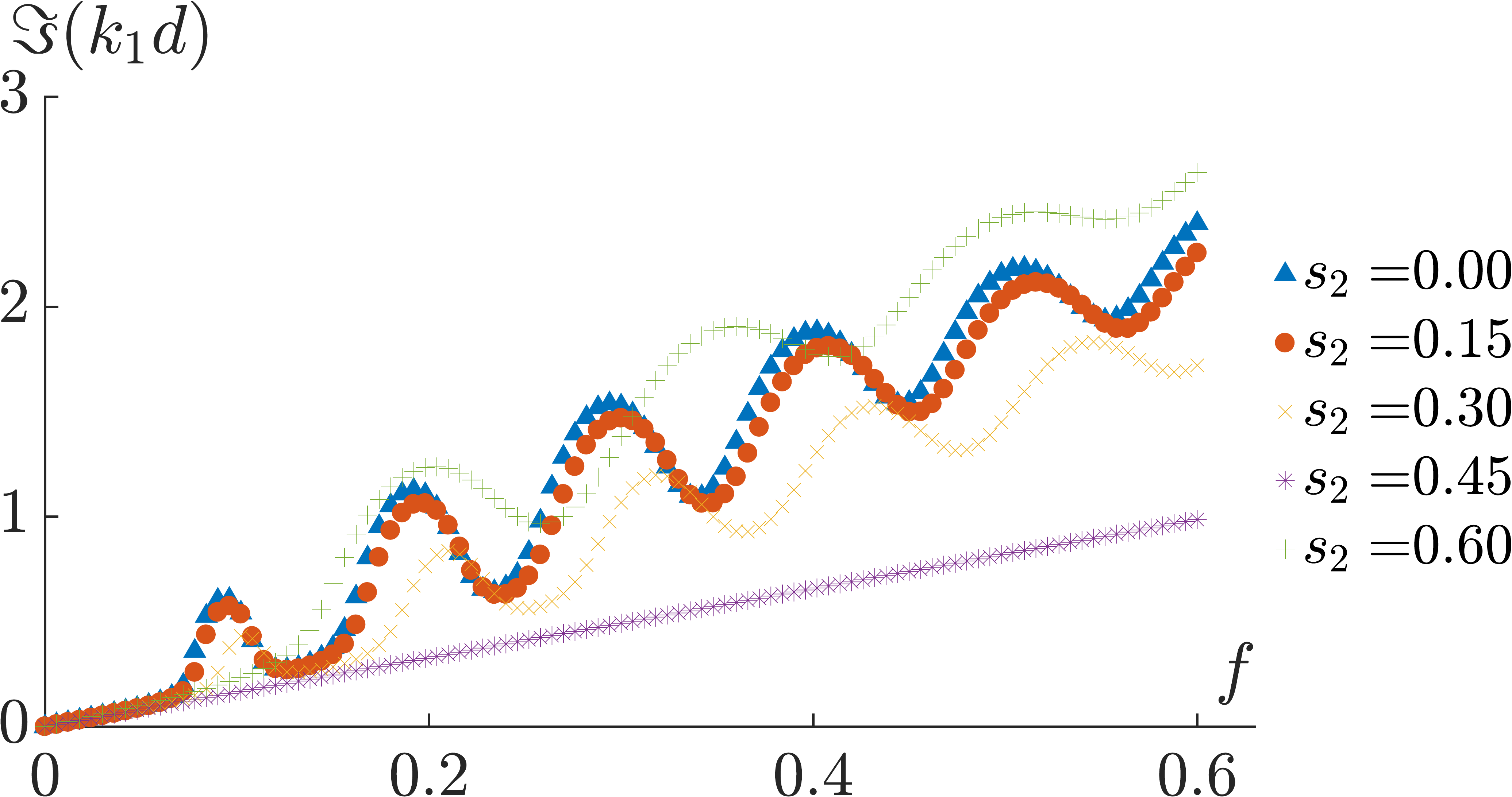}
		\caption{\label{fig:Im-k1d}}
	\end{subfigure}
	\begin{subfigure}[b]{0.5\linewidth}
		\centering\includegraphics[height=95pt]{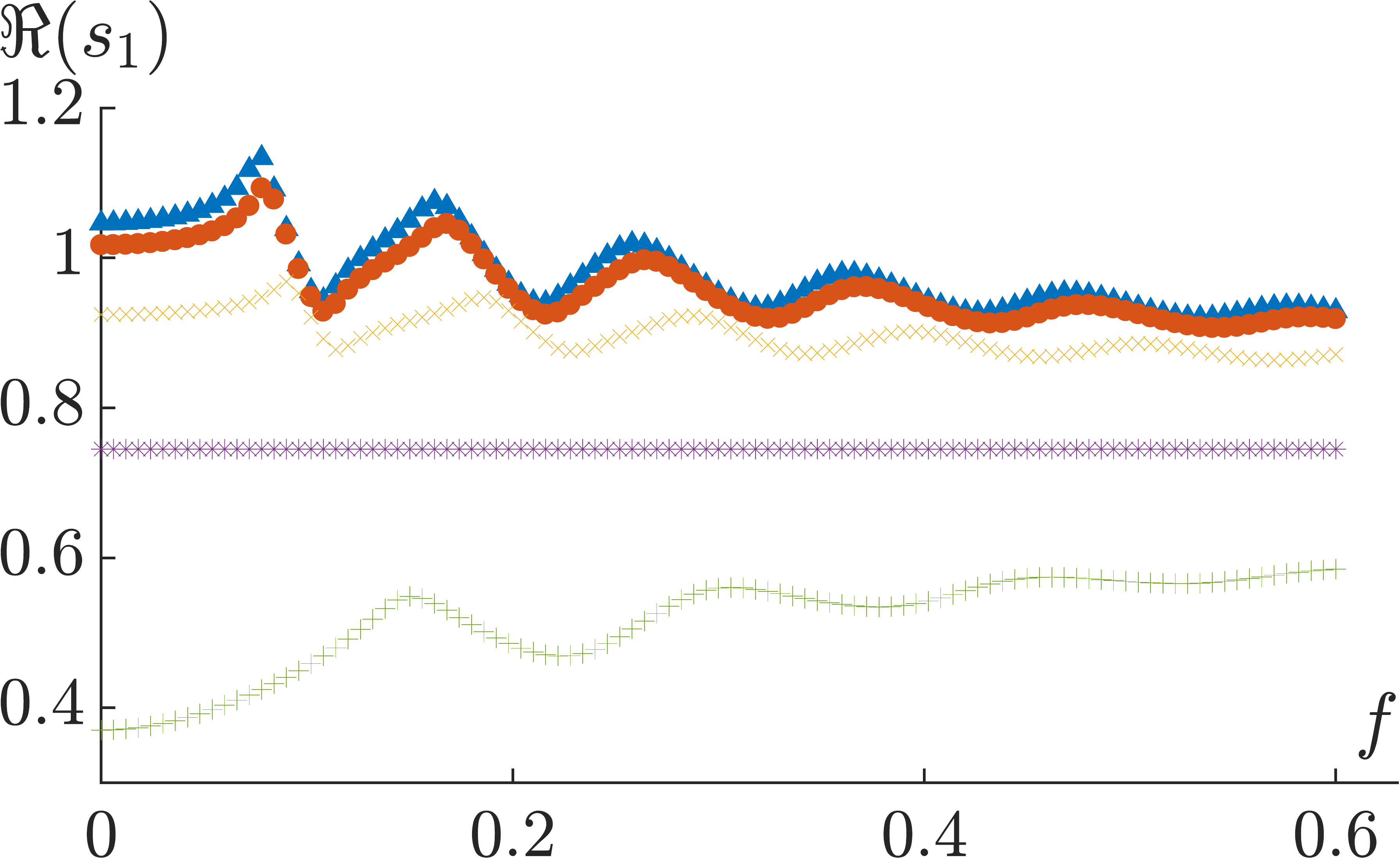}
		\caption{\label{fig:Re-s1vf}}
	\end{subfigure}%
	\begin{subfigure}[b]{0.5\linewidth}
		\centering\includegraphics[height=95pt]{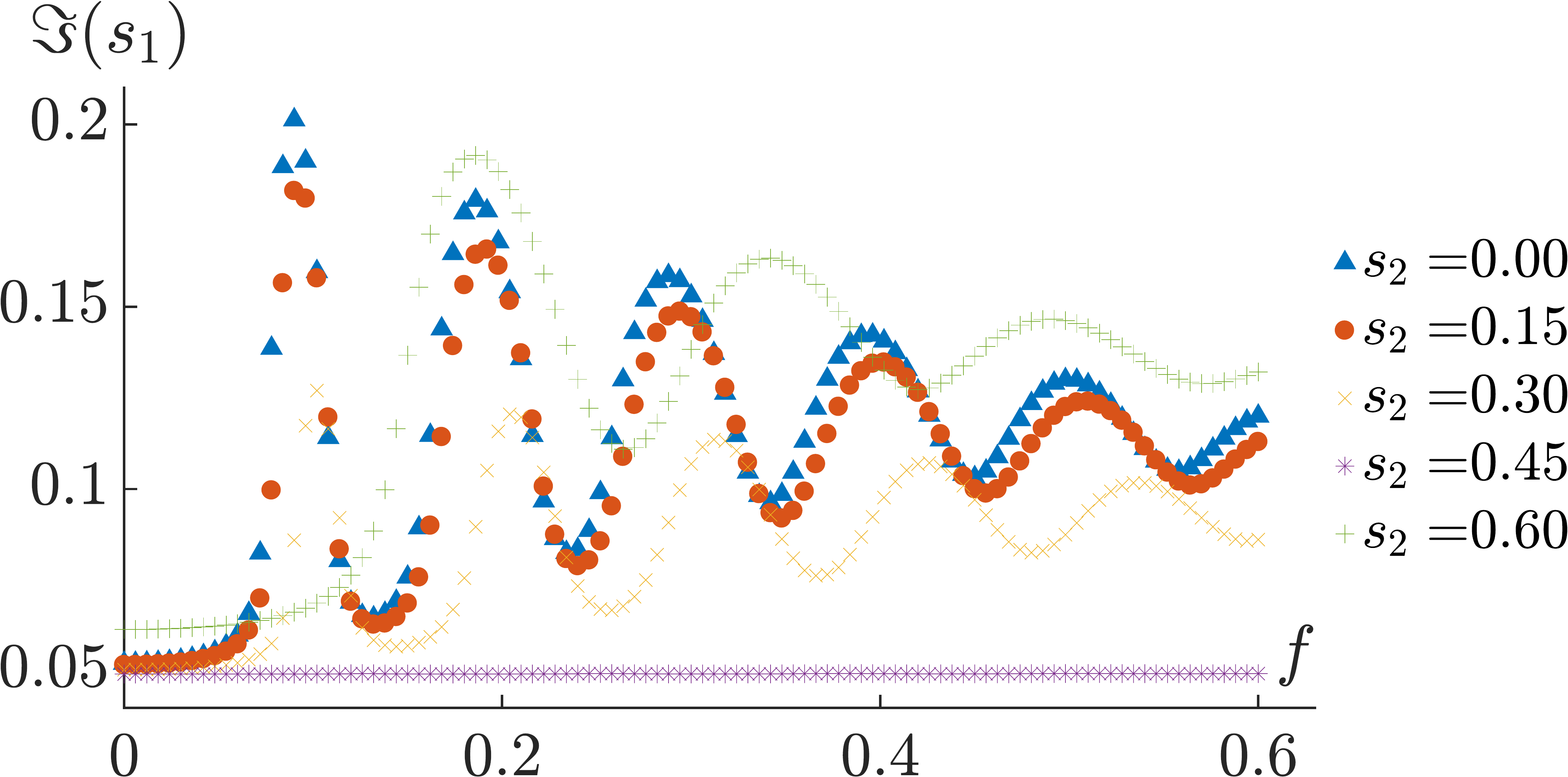}
		\caption{\label{fig:Im-s1vf}}
	\end{subfigure}
		\caption{\label{fig:k1d}(\subref{fig:Re-k1d}) Real and  (\subref{fig:Im-k1d}) imaginary parts of the normalized wave number $k_1d$, as functions of frequency, $f$ for different values of $s_2$.  Due to the symmetry of the system, the other solution would be the negative of the presented values. (\subref{fig:Re-s1vf}) and (\subref{fig:Im-s1vf}) are the real and imaginary parts of the slowness component $s_1 = k_1/\om$, respectively. The dependence of $s_1$ on frequency demonstrates nonlocal physics, as otherwise the slowness will be constant in frequency. The only such special case is happening at $s_2=0.45$. The phase ambiguity is removed through maintaining continuity of $k_1d$ by adding $2\pi$ when necessary.}
\end{figure}

\begin{figure}[!ht]
	\centering
	\begin{subfigure}[b]{0.5\linewidth}
		\centering\includegraphics[height=100pt,left]{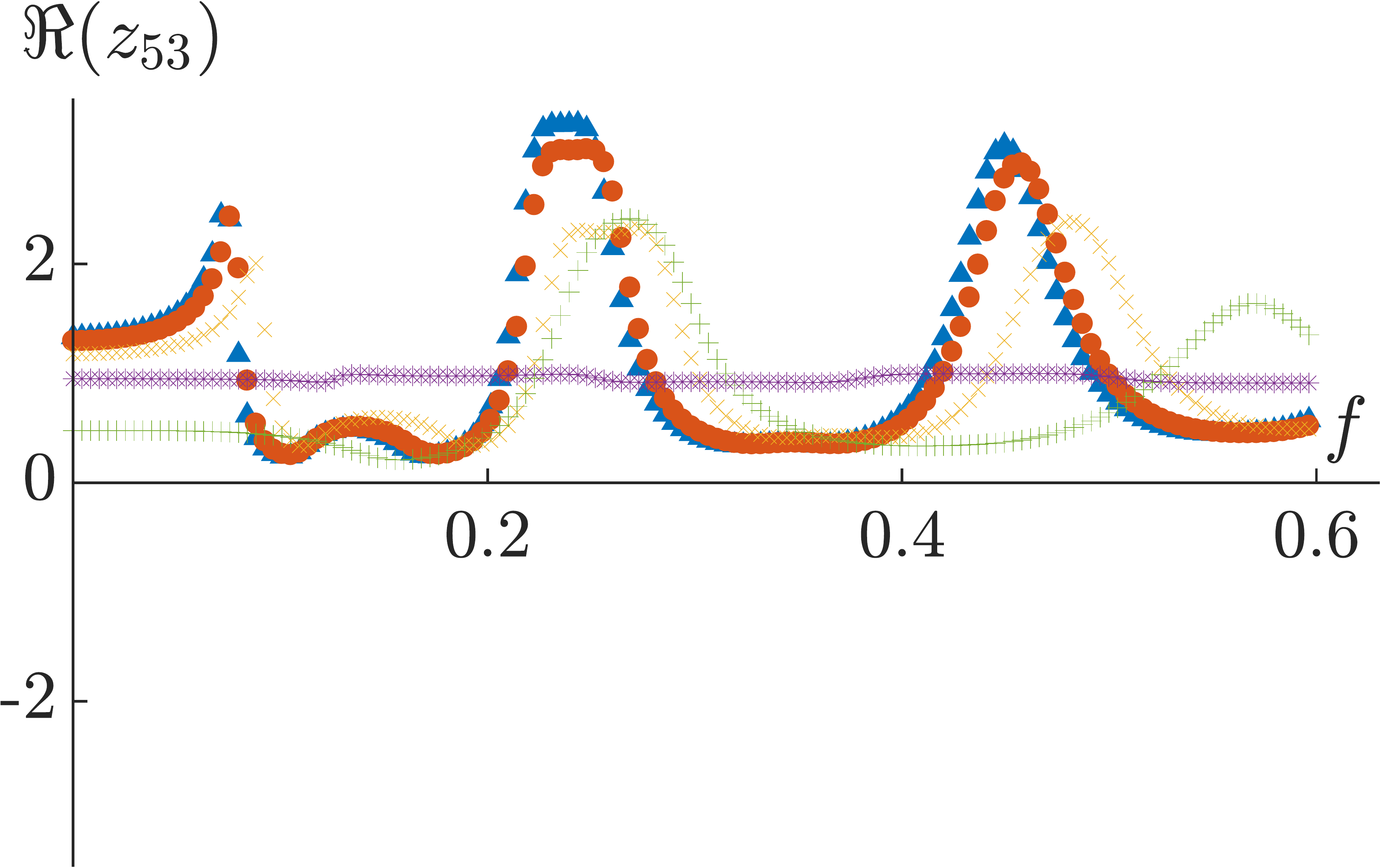}
		\caption{\label{fig:Re-z53}}
	\end{subfigure}%
	\begin{subfigure}[b]{0.5\linewidth}
		\centering\includegraphics[height=100pt,left]{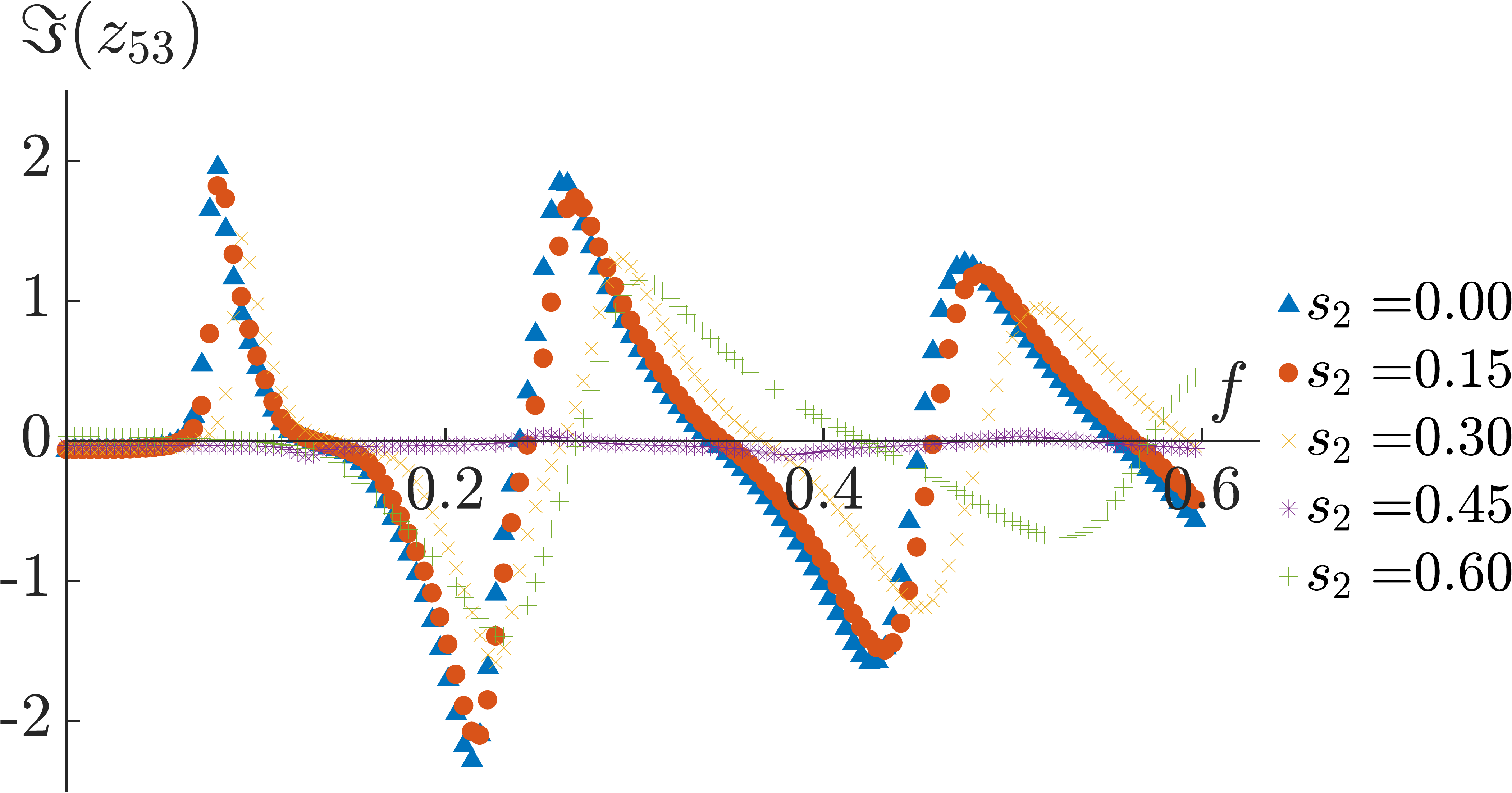}
		\caption{\label{fig:Im-z53}}
	\end{subfigure}
	\caption{\label{fig:z53}(\subref{fig:Re-z53}) and (\subref{fig:Im-z53}) show the real and imaginary parts of the impedance, $z_{53}$, as functions of frequency, $f$, for different values of $s_2$, respectively. Again, the impedance appears as independent of frequency only at $s_2=0.45$.}
\end{figure}

Here, the proposed method is applied to an example of a symmetric system in order to expand the determination of the constitutive parameters into the stop bands and upper pass bands. The periodic layered medium of interest here is [PMMA, Brass, PMMA] with $(d_j) = (2.5, 0.2, 2.5)$, $(\rho_j) = (1, 8, 1)$, and $(\mu_j) = (1.2, 40, 1.2)$  which is shown in Figure~(\ref{fig:schcomp}). The PMMA phase is modeled as a lossy material with 5\% loss in terms of wave speed, i.e. ${c_1''/c_1' = c_3''/c_3' = -0.05}$ (${\approx10\%}$ loss tangent in modulus).
In the following, the unit system [mm, $\mu$s, mg] for length, time, and mass are used which results in [km/s, GPa, g/cm$^{3}$, MHz, MRayl] units for velocity, stress, density, frequency, and impedance, respectively. The complex reflection and transmission coefficients into PC are shown in Figure~(\ref{fig:Scat}) for a single unit cell. $\theta=0, \pi/6$ graphs for a five unit cell slab are also added in order to observe how increasing the number of unit cells can affect the scattering results.
\paragraph{Dispersion, wave vector, and impedance} Figures~(\ref{fig:k1d}) and (\ref{fig:z53}) depict the dispersion results extracted from the scattering data based on Equation~(\ref{eq:STM}). Note that the results are independent of the number of unit cells in the slab $n_c$. Furthermore, using the eigenvalue method in Equation~\eqref{eq:eign} also gives the same exact wave vector component $k_1$ and impedance $z_{53}$. Here the normalized phase advance, $k_1 d$, is shown as a function of frequency, $f$, for different values of $s_2$. The phase ambiguity is removed by adding $2\pi$ whenever needed to maintain continuity, leading to a single positive phase advance solution $0 \leq \Re(k_1 d) \leq \infty$. Due to the symmetry of the structure, the second solution is simply the negative of this result. In general, for higher values of ${s_2 = k_2/\omega}$ the stop bands, associated with peaks in $\Im(k_1d)$, become wider. However, for the incident angle ${\theta=\pi/6}$, the PC material constants requires $s_2 = 0.45$, for which the scattering calculation leads to consistently high amplitude of transmission and low reflection. It is also observed that at this value of $s_2$, $k_1 d$ phase advance is almost exactly a linear function of frequency, essentially rendering the medium non-dispersive, i.e leading to constant $s_1$ and $z_{53}$ in terms of frequency. 

\begin{figure}[!ht]
	\centering
	\begin{subfigure}[b]{0.5\linewidth}
		\centering\includegraphics[height=105pt,center]{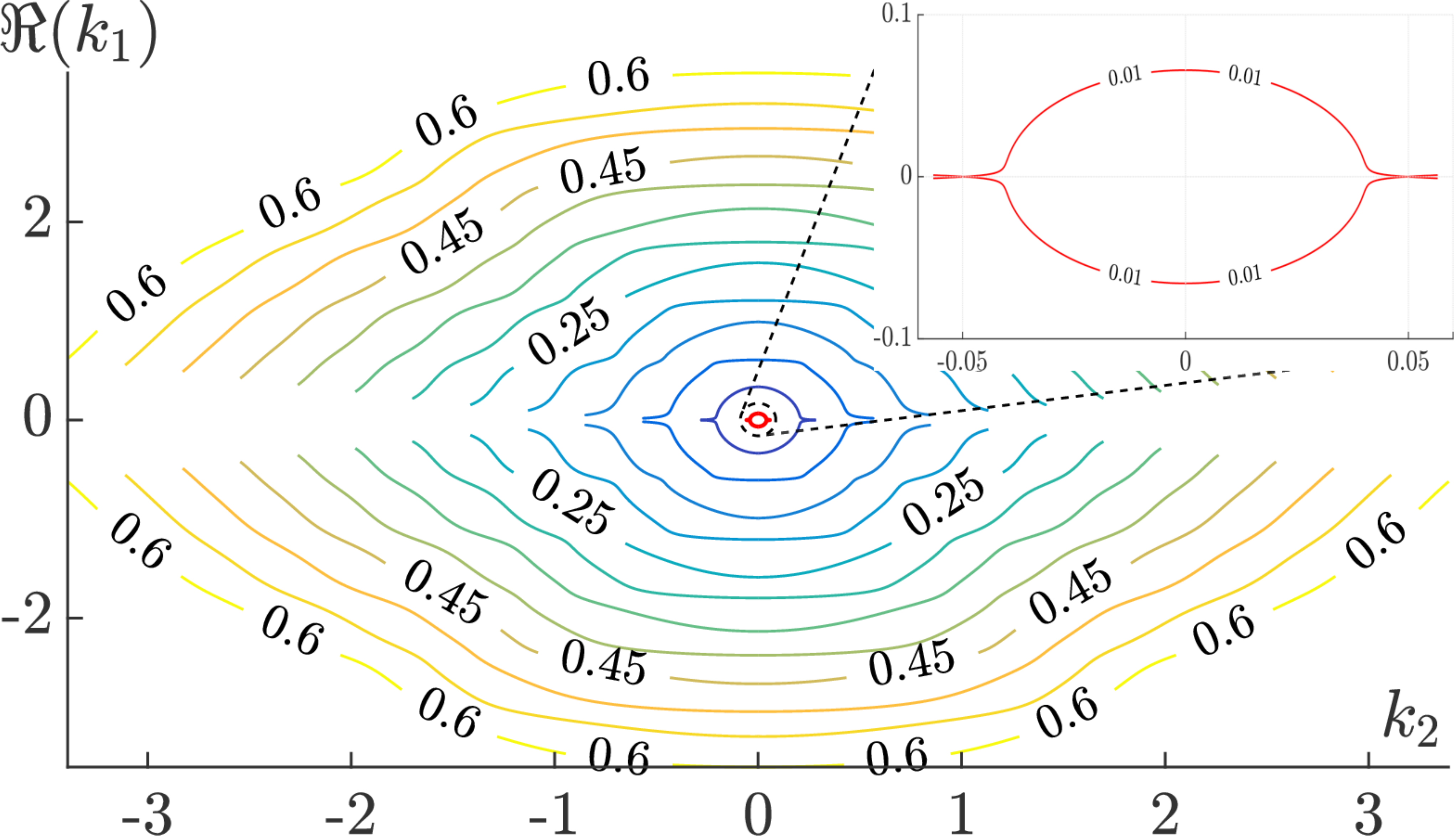}
		\caption{\label{fig:Re-k1c}}
	\end{subfigure}%
	\begin{subfigure}[b]{0.5\linewidth}
		\centering\includegraphics[height=105pt,center]{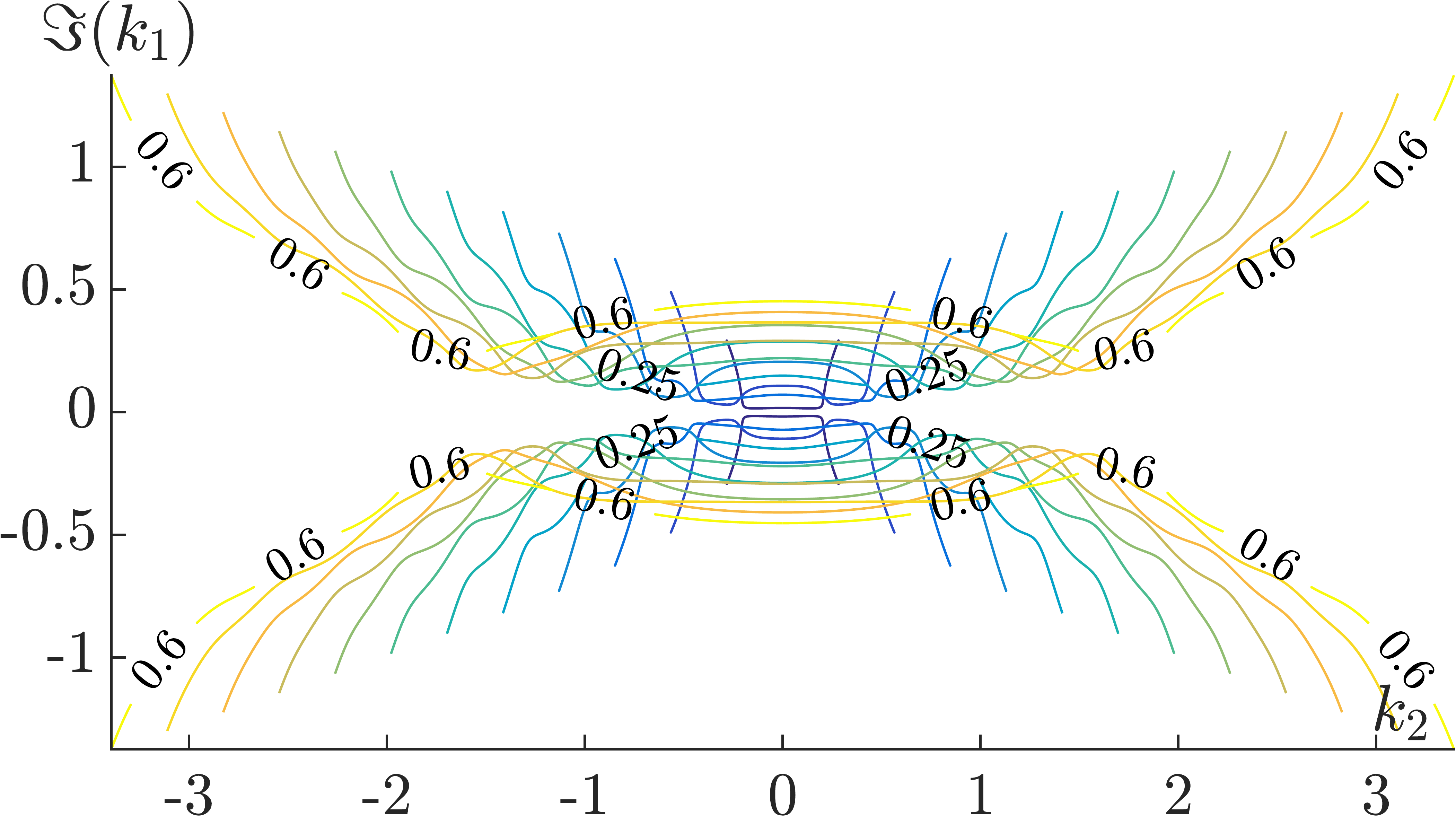}
		\caption{\label{fig:Im-k1c}}
	\end{subfigure}
	\begin{subfigure}[b]{0.5\linewidth}
		\centering\includegraphics[height=102pt,center]{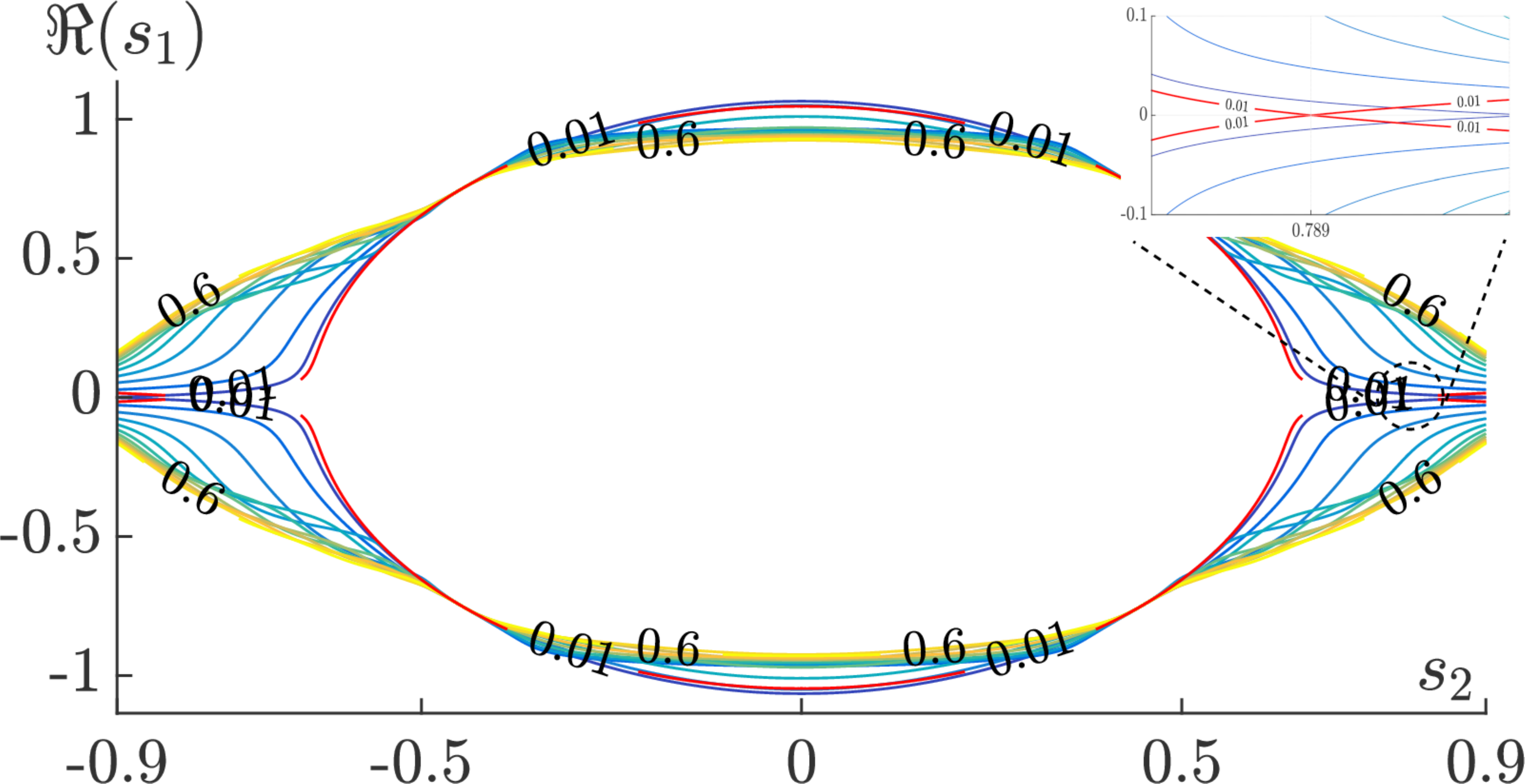}
		\caption{\label{fig:Re-s1c}}
	\end{subfigure}%
	\begin{subfigure}[b]{0.5\linewidth}
		\centering\includegraphics[height=102pt,center]{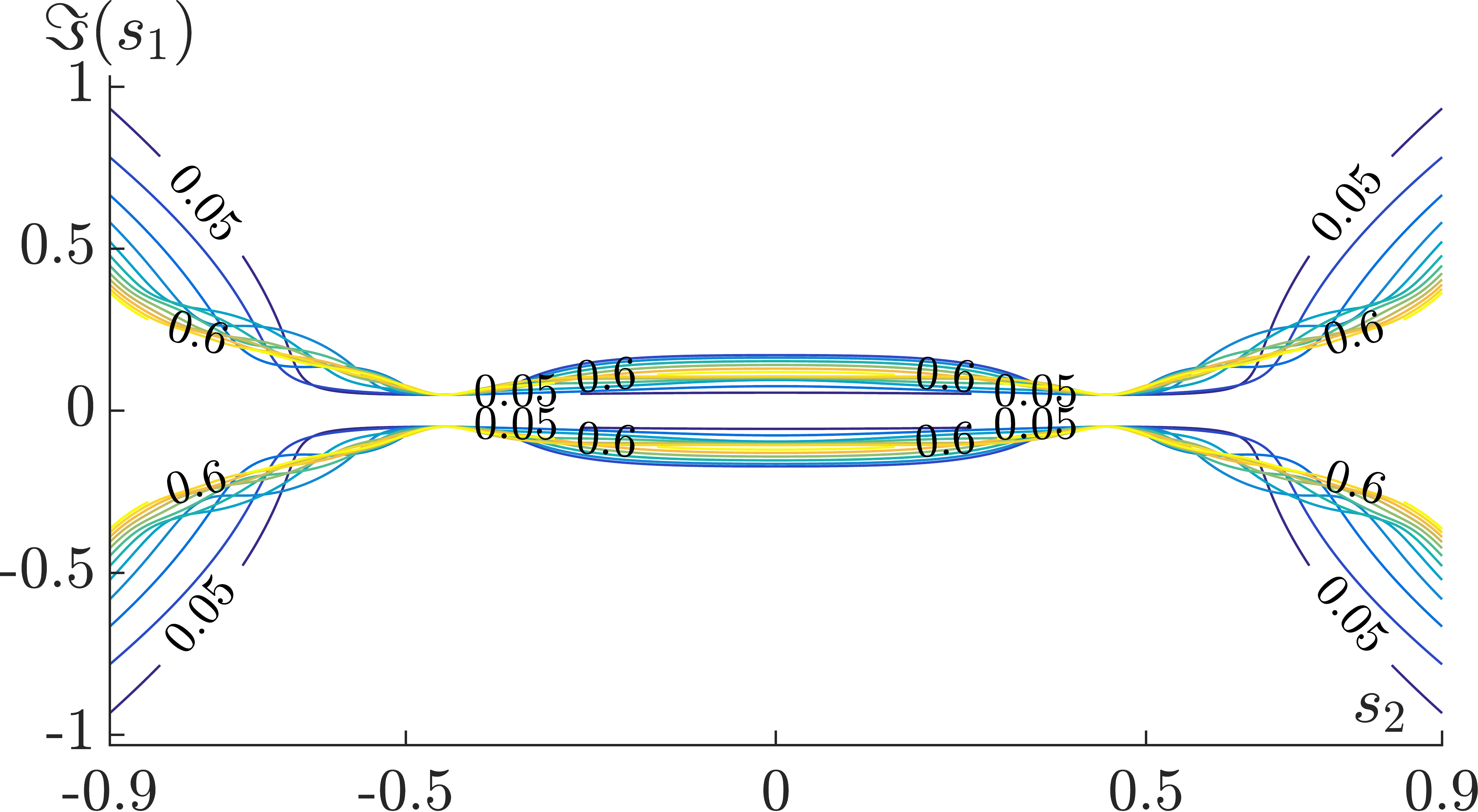}
		\caption{\label{fig:Im-s1c}}
	\end{subfigure}
		\caption{\label{fig:s1k1c}(\subref{fig:Re-k1c}) and (\subref{fig:Im-k1c}) are the contour plot representations of the dispersion surfaces for real and imaginary parts of $k_1$ vs. $k_2$ and frequency (contour values). (\subref{fig:Re-s1c}) and (\subref{fig:Im-s1c}) show isofrequency contours in the 2D spaces of $s_2$ and real and imaginary parts of $s_1$, respectively. In (\subref{fig:Re-s1c}) it is shown that for $f=0.01$, $\Re(s_1)$ is changing sign at $s_2=\pm 0.789$ which results in $\Re(k_1)$ sign change at $k_2 = \pm 0.05$, (\subref{fig:Re-k1c}). In a spatially non-dispersive system, the slowness graphs will be independent of frequency, therefore this graphs will only show a single curve for SH waves. Note that all different contours converge near $s_2 = 0.45$. In other words, the system is nearly non-dispersive in the neighborhood of this point $\bm{s} \approx (\pm (0.7497 + 0.0493i), \pm 0.4470)^T$. At this point $z_{53} = 0.9563 - 0.0328i$.} 
\end{figure}
Figure~(\ref{fig:s1k1c}) shows two representations of the dispersion surfaces. One can observe that for high amplitudes of $|s_2| > 0.5$,  $\Im(s_1)$ increases significantly. The sign of $\Re(s_1)$ is changing for $|s_2| > 0.789$; see the insets.
The isofrequency contours appear to converge around  $\bm{s} \approx (\pm( 0.750 + 0.049i), \pm 0.447)^T$. This is where the system appears non-dispersive.
The value of impedance around this point is also nearly independent of frequency $z_{53}=0.9563 - 0.0328i$.

\begin{figure}[!ht]
	\centering\includegraphics[height=120pt]{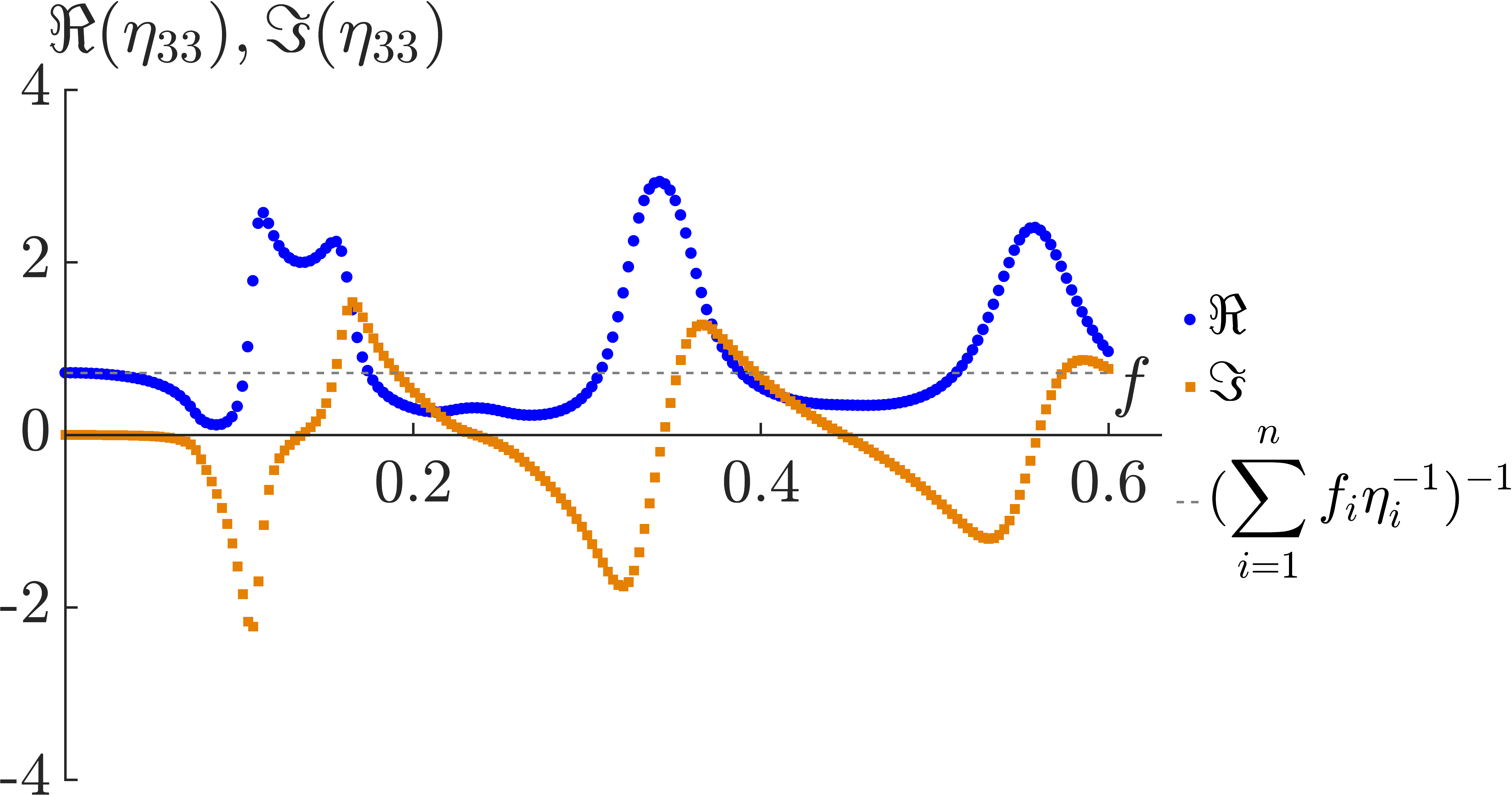}
	\caption{\label{fig:eta33} The real and imaginary parts of $\eta_{33}$, as functions of frequency, $f$, are shown here. The overall $\eta_{33}$ would be equal to the Reuss average of component layers at very low frequencies.}
\end{figure}

\paragraph{Constitutive functions} In both approaches discussed above, the inverse density quantity, $\eta_{33}$, may be chosen to be independent of wave vector or slowness $s_2$ (Equations~\eqref{eq:eta33sc1},\eqref{eq:eta33sc2} which give the same numerical value). The results are shown in Figure~(\ref{fig:eta33}). At very low frequencies, the overall density of the media should match a simple volume average, which translates into the Reuss average formula for $\eta_{33}$. This is shown in Fig~(\ref{fig:eta33}), where $\Re(\eta_{33})$ is tangent to this value and $\Im(\eta_{33}) \rightarrow 0$ at low frequencies. 

\begin{figure}[!ht]
	\centering
	\begin{subfigure}[b]{0.5\linewidth}
		\centering\includegraphics[height=100pt,left]{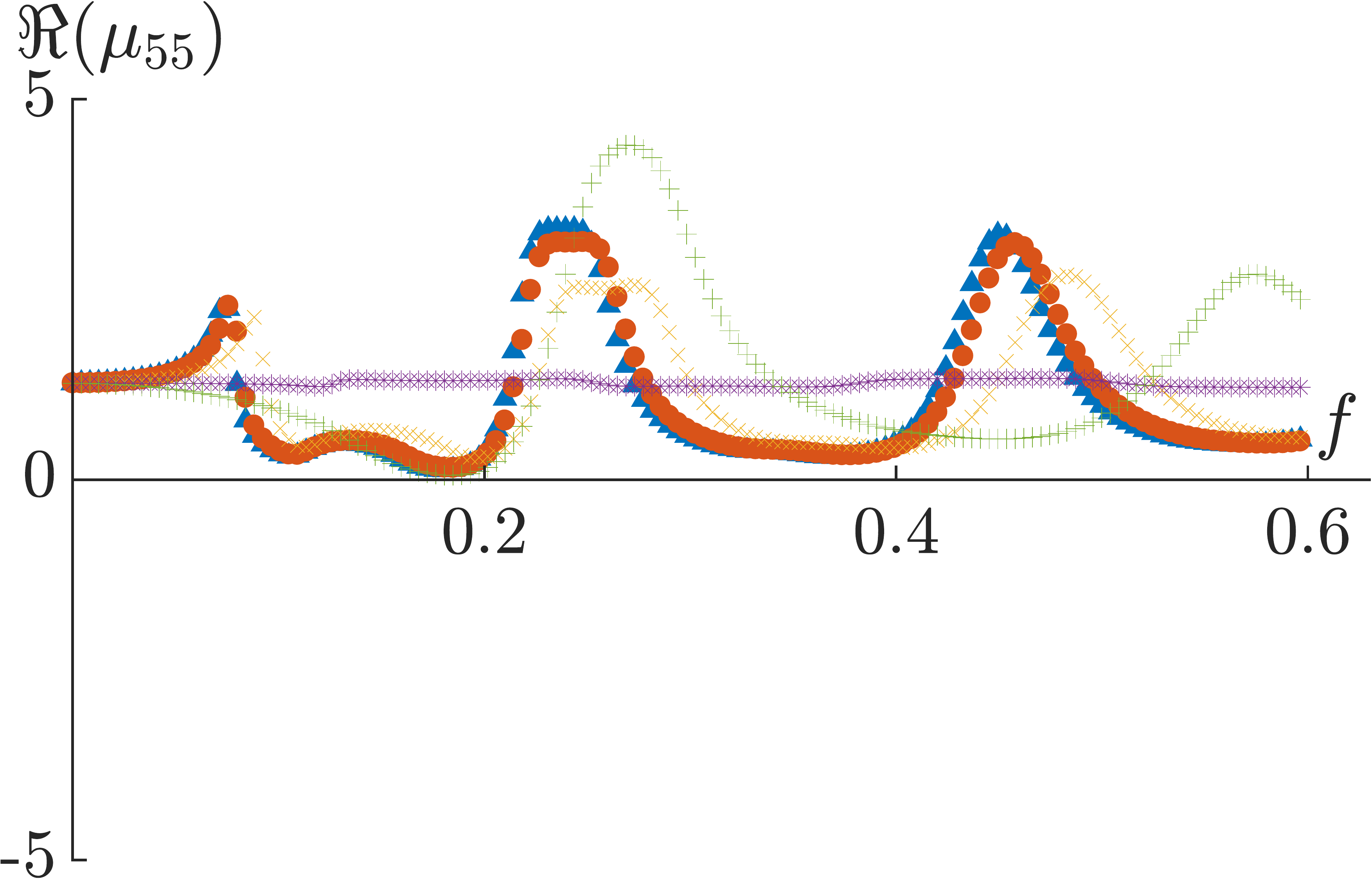}
		\caption{\label{fig:Re-m55sc1}}
	\end{subfigure}%
	\begin{subfigure}[b]{0.5\linewidth}
		\centering\includegraphics[height=100pt,left]{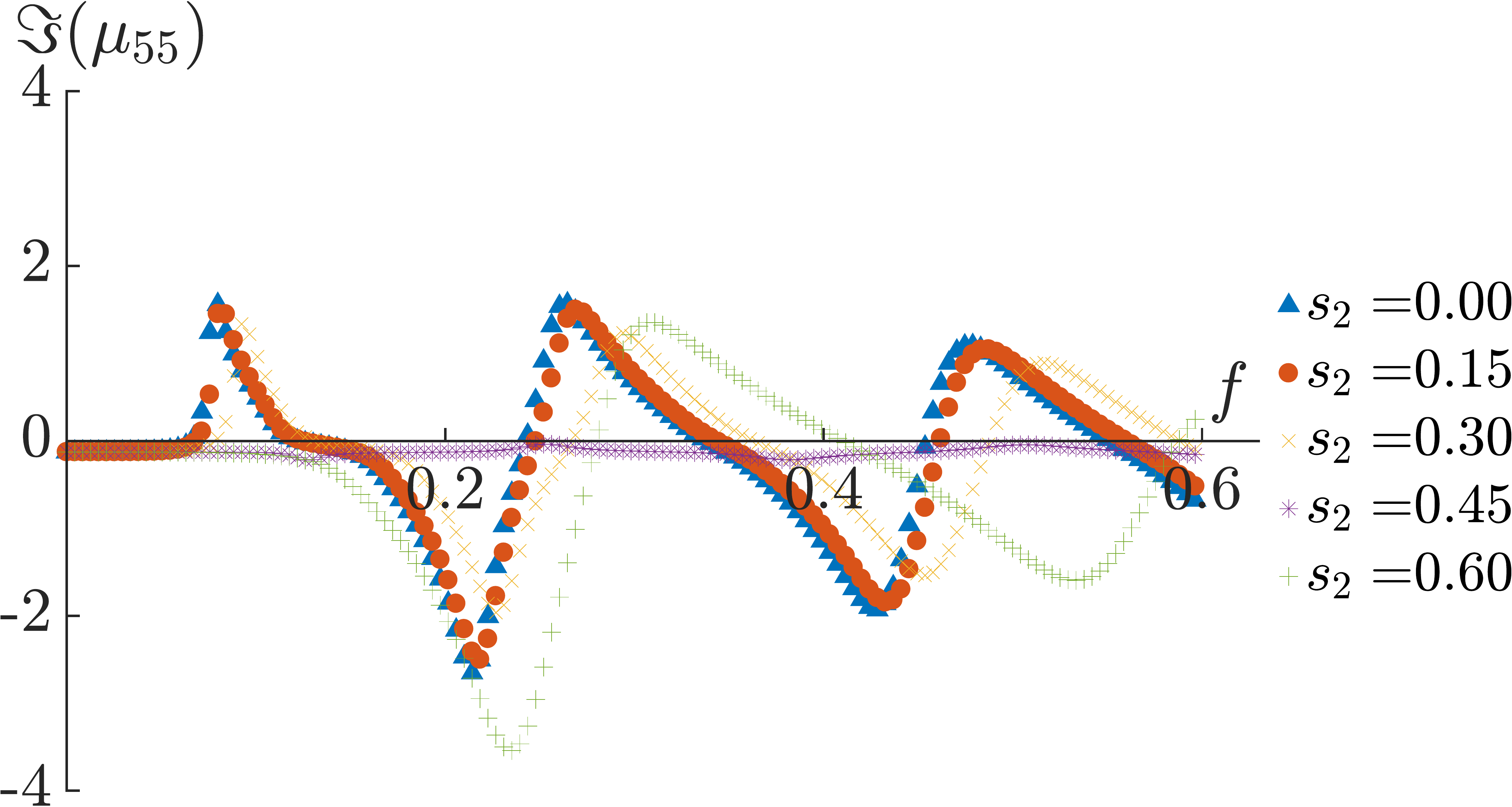}
		\caption{\label{fig:Im-mu55sc1}}
	\end{subfigure}
	\caption{\label{fig:mu55}(\subref{fig:Re-m55sc1}) and (\subref{fig:Im-mu55sc1}) show the real and imaginary parts of the  $\mu_{55}$ as functions of frequency, $f$, for different values of $s_2$, respectively, when a diagonal constitutive tensor is used.}
\end{figure}

\begin{figure}[!ht]
	\centering
	\begin{subfigure}[b]{0.5\linewidth}
		\centering\includegraphics[height=100pt,left]{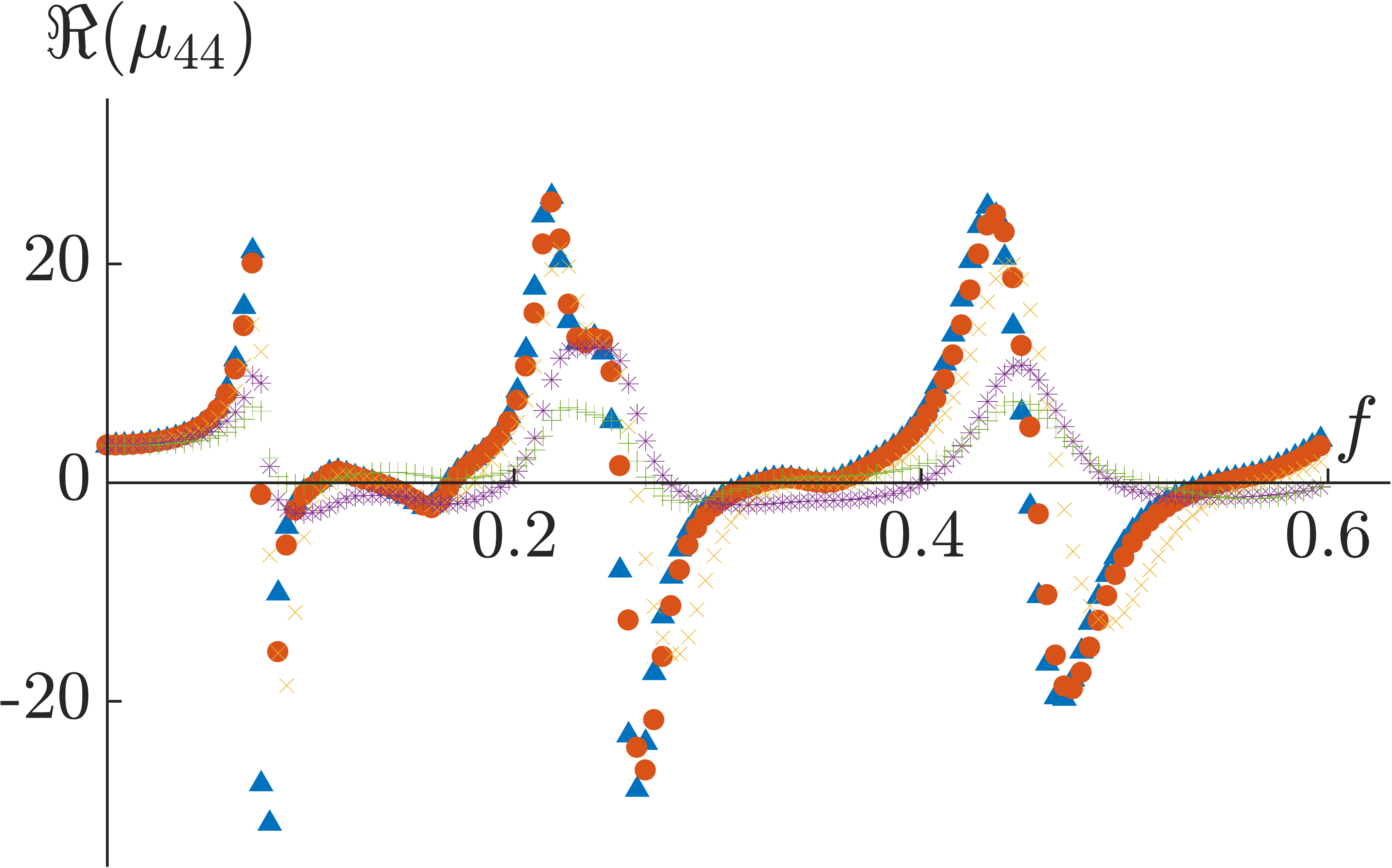}
		\caption{\label{fig:Re-m44sc1}}
	\end{subfigure}%
	\begin{subfigure}[b]{0.5\linewidth}
		\centering\includegraphics[height=100pt,left]{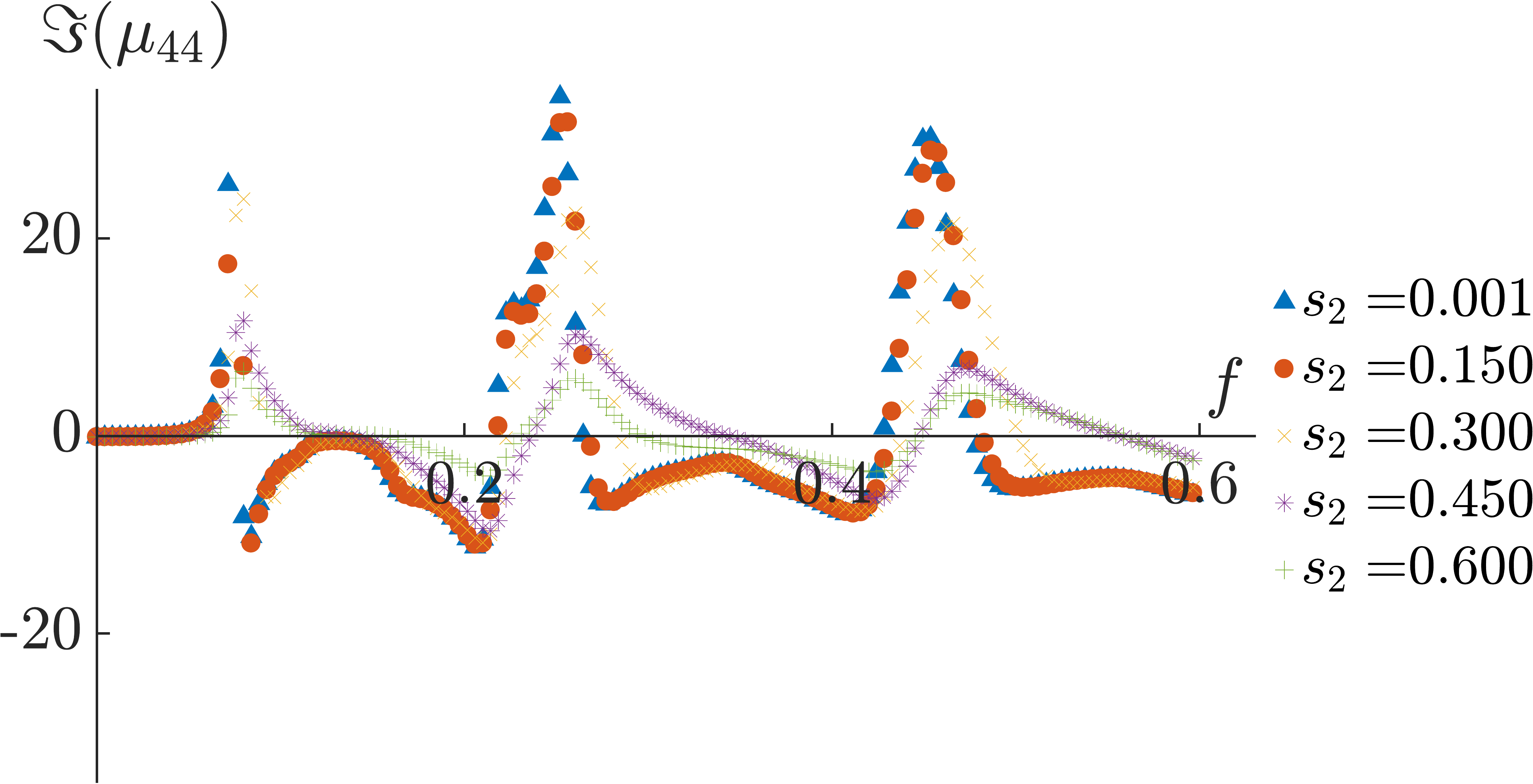}
		\caption{\label{fig:Im-mu44sc1}}
	\end{subfigure}
	\caption{\label{fig:mu44}(\subref{fig:Re-m44sc1}) and (\subref{fig:Im-mu44sc1}) show the real and imaginary parts of $\mu_{44}$ as functions of frequency, $f$, for different values of $s_2$, respectively, when a diagonal constitutive tensor is used.}
\end{figure}

For a diagonal constitutive tensors, the stiffness quantities will have to be functions of the wave vector. In this case Equations~\eqref{eq:m55dd} and \eqref{eq:mu44sc1} will be used. As it can be seen from Figure~(\ref{fig:mu55}), the value of $\mu_{55}$ is a strong function of propagation direction and even becomes relatively constant around $s_2 = 0.45$. $\mu_{44}$ is shown in Figure~(\ref{fig:mu44}) and also demonstrates significant non-locality and sensitivity to the wave vector, especially in and around stop bands.\footnote{$\mu_{44}$ is not calculated for $s_2=0$ (Equation~\ref{eq:mu44sc1}), and therefore is shown starting from $s_2=0.001$.}

\begin{figure}[!ht]
	\centering
	\begin{subfigure}[b]{0.5\linewidth}
		\centering\includegraphics[height=100pt,left]{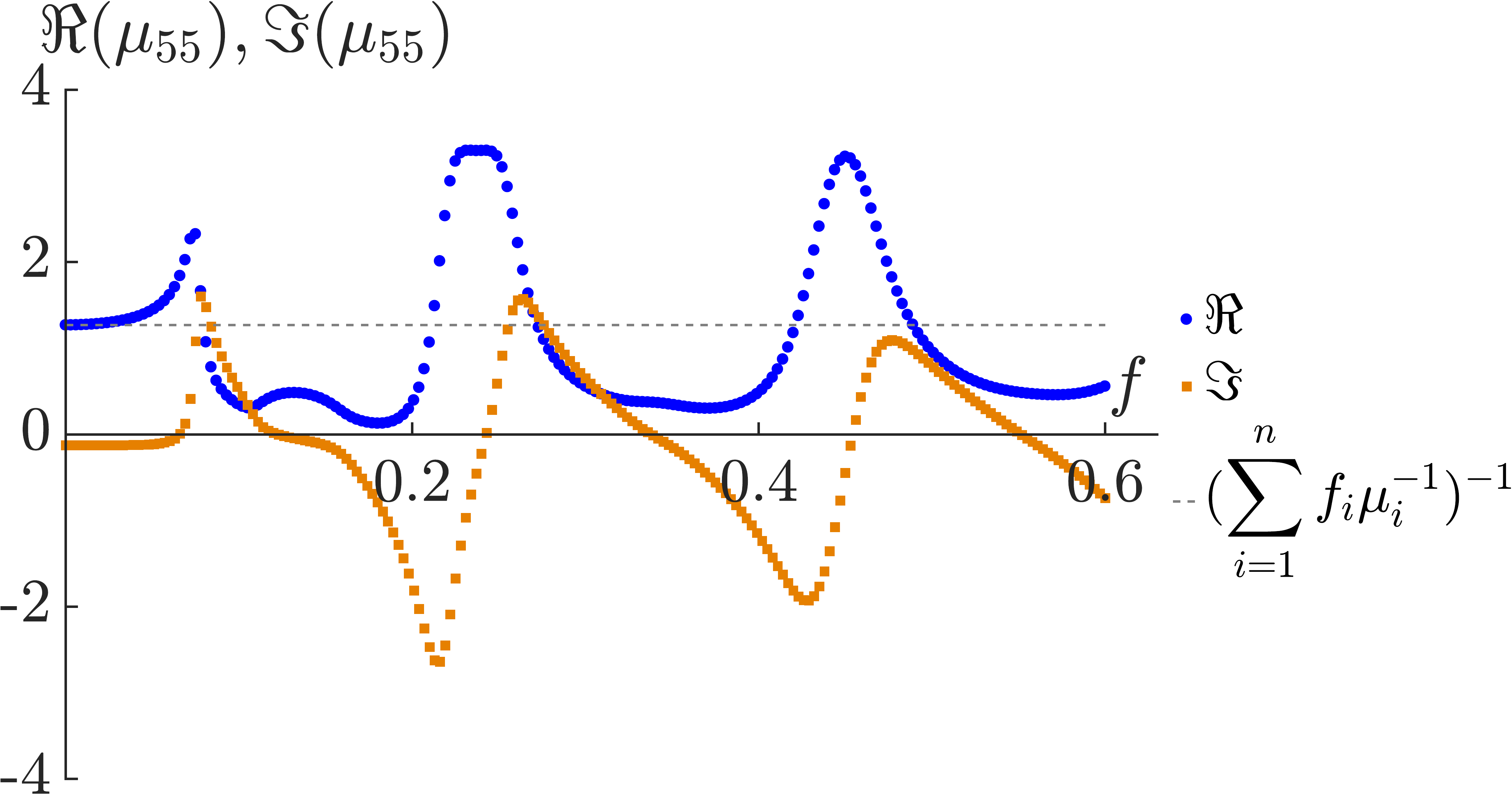}
		\caption{\label{fig:mu55sc2}}
	\end{subfigure}%
	\begin{subfigure}[b]{0.5\linewidth}
		\centering\includegraphics[height=100pt,left]{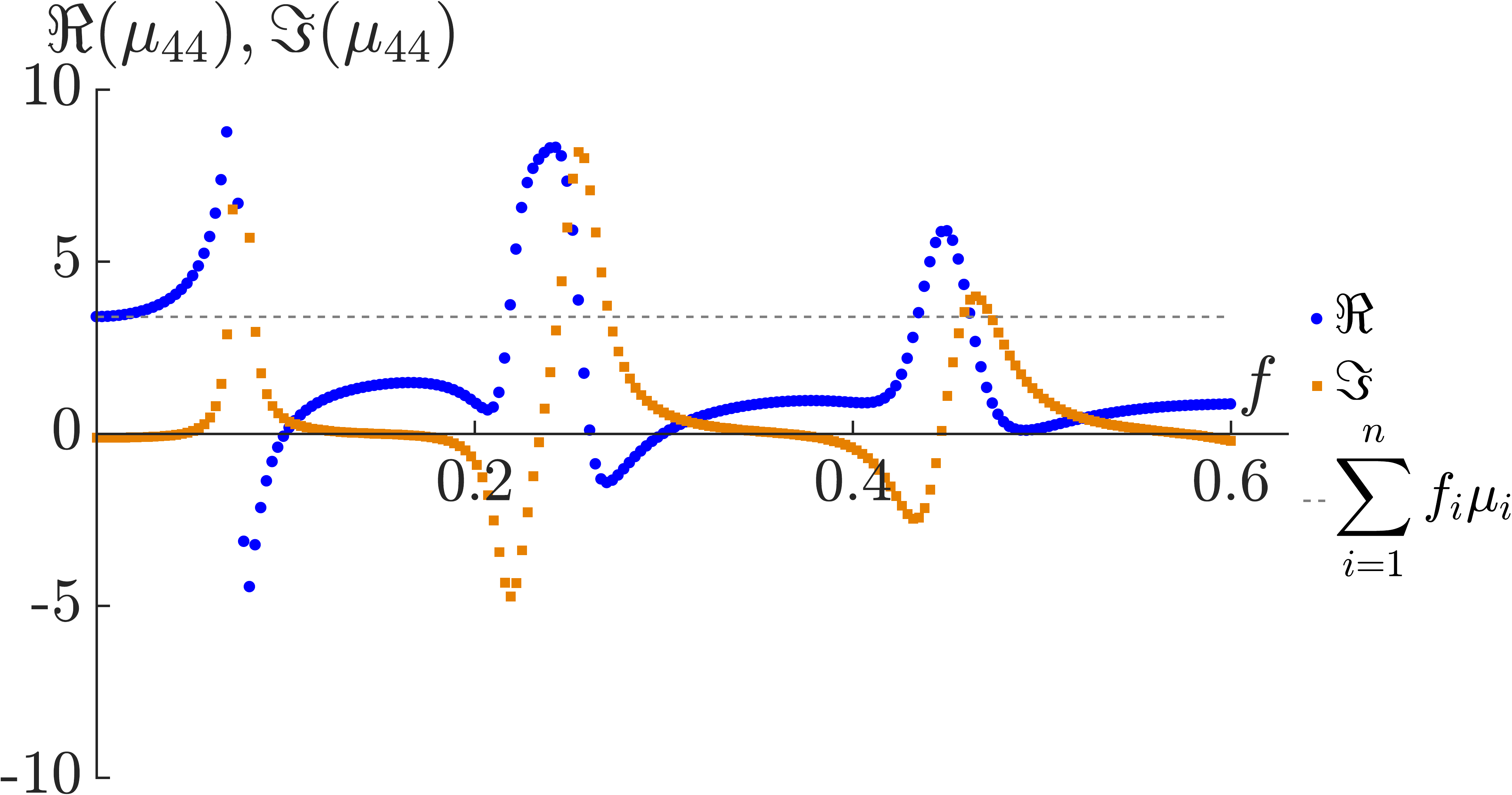}
		\caption{\label{fig:mu44sc2}}
	\end{subfigure}
	\caption{\label{fig:mu55an44}(\subref{fig:mu55sc2}) and (\subref{fig:mu44sc2}) show the real and imaginary parts of the moduli $\mu_{55}$ and $\mu_{44}$ as functions of frequency, $f$, when they are considered independent of the wave vector. It should be noted that in this case, Reuss and Voigt averages provide accurate estimates of overall, $\mu_{55}$ and $\mu_{44}$, respectively at very low frequencies as one expects for quasi-static situations.}
\end{figure}

\begin{figure}[!ht]
	\centering
	\begin{subfigure}[b]{0.5\linewidth}
		\centering\includegraphics[height=100pt,left]{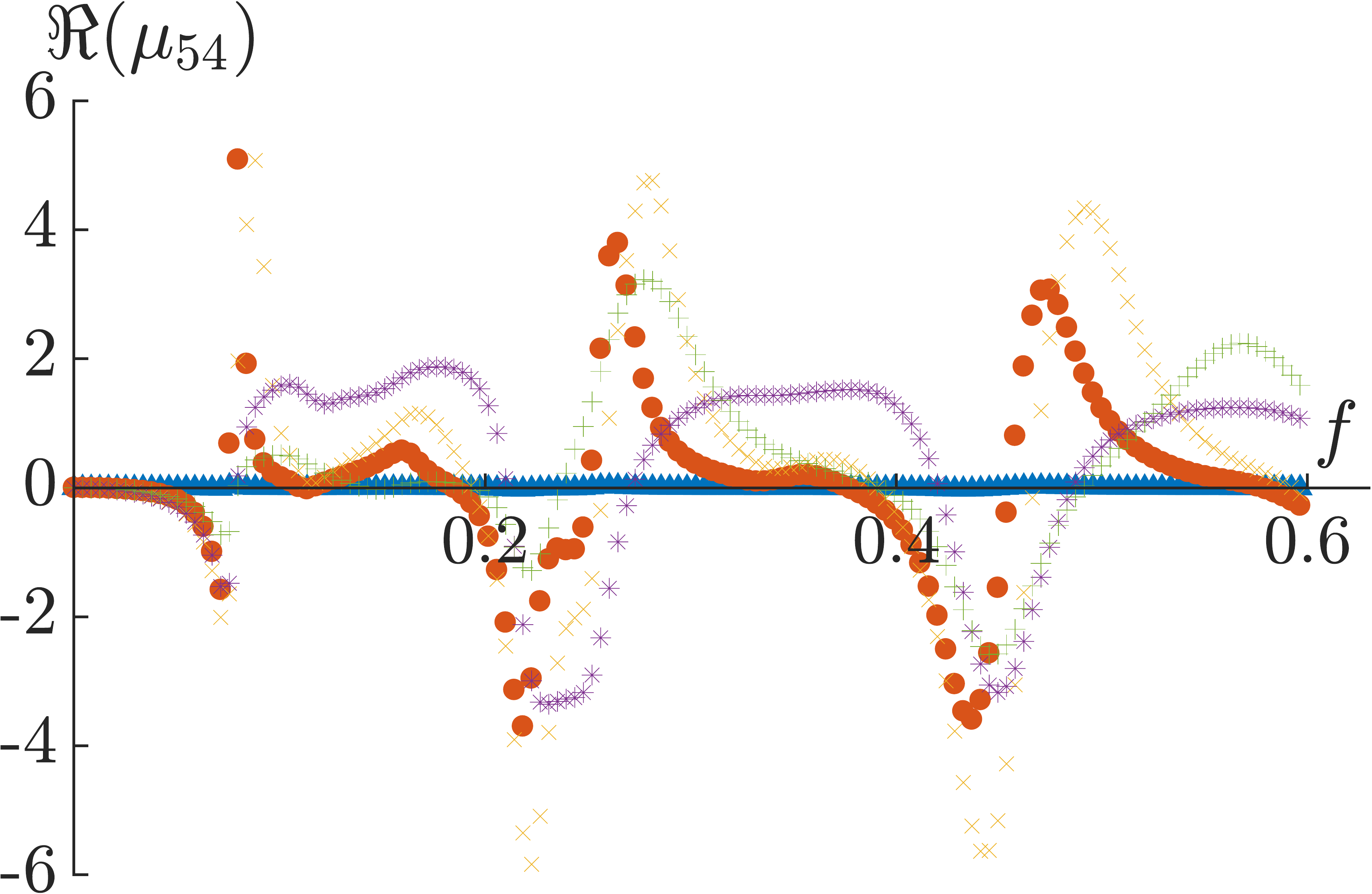}
		\caption{\label{fig:Re-mu54}}
	\end{subfigure}%
	\begin{subfigure}[b]{0.5\linewidth}
		\centering\includegraphics[height=100pt,left]{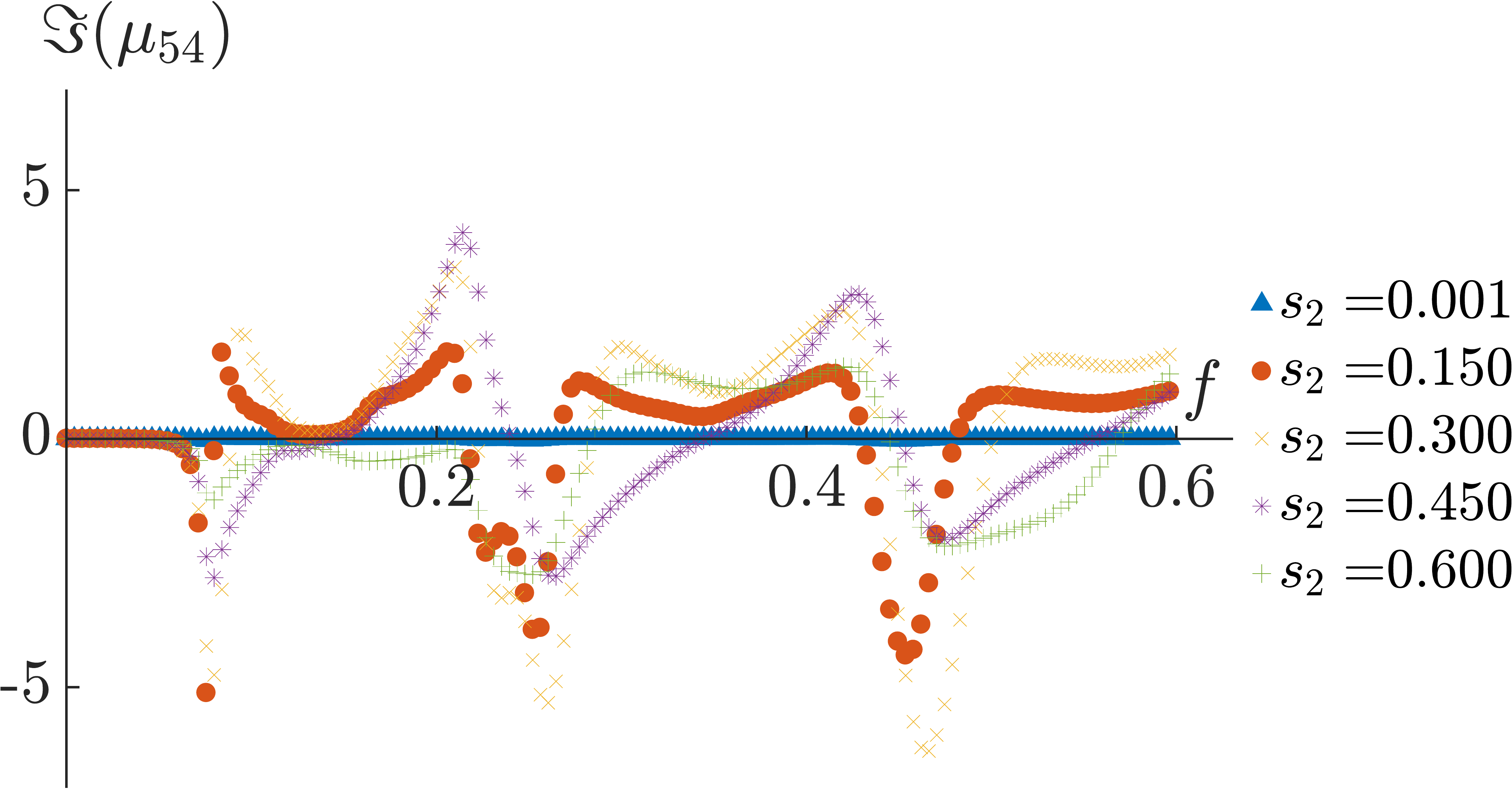}
		\caption{\label{fig:Im-mu54}}
	\end{subfigure}
	\caption{\label{fig:mu54}(\subref{fig:Re-mu54}) and (\subref{fig:Im-mu54}) show the real and imaginary parts of the moduli, $\mu_{54}$ as functions of frequency, $f$.}
\end{figure}

\begin{figure}[!ht]
	\centering
	\begin{subfigure}[b]{0.5\linewidth}
		\centering\includegraphics[height=100pt,left]{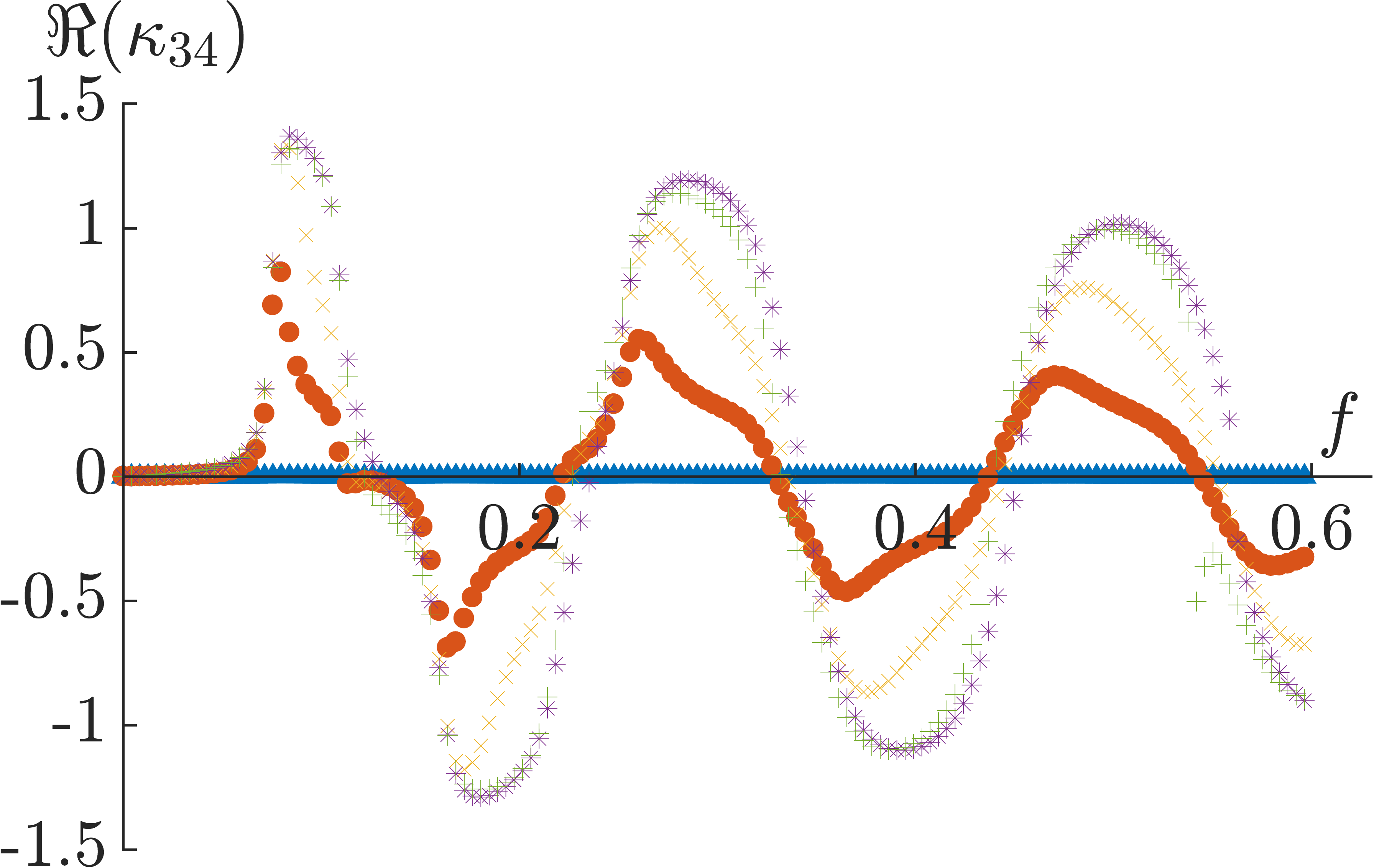}
		\caption{\label{fig:Re-ka34}}
	\end{subfigure}%
	\begin{subfigure}[b]{0.5\linewidth}
		\centering\includegraphics[height=100pt,left]{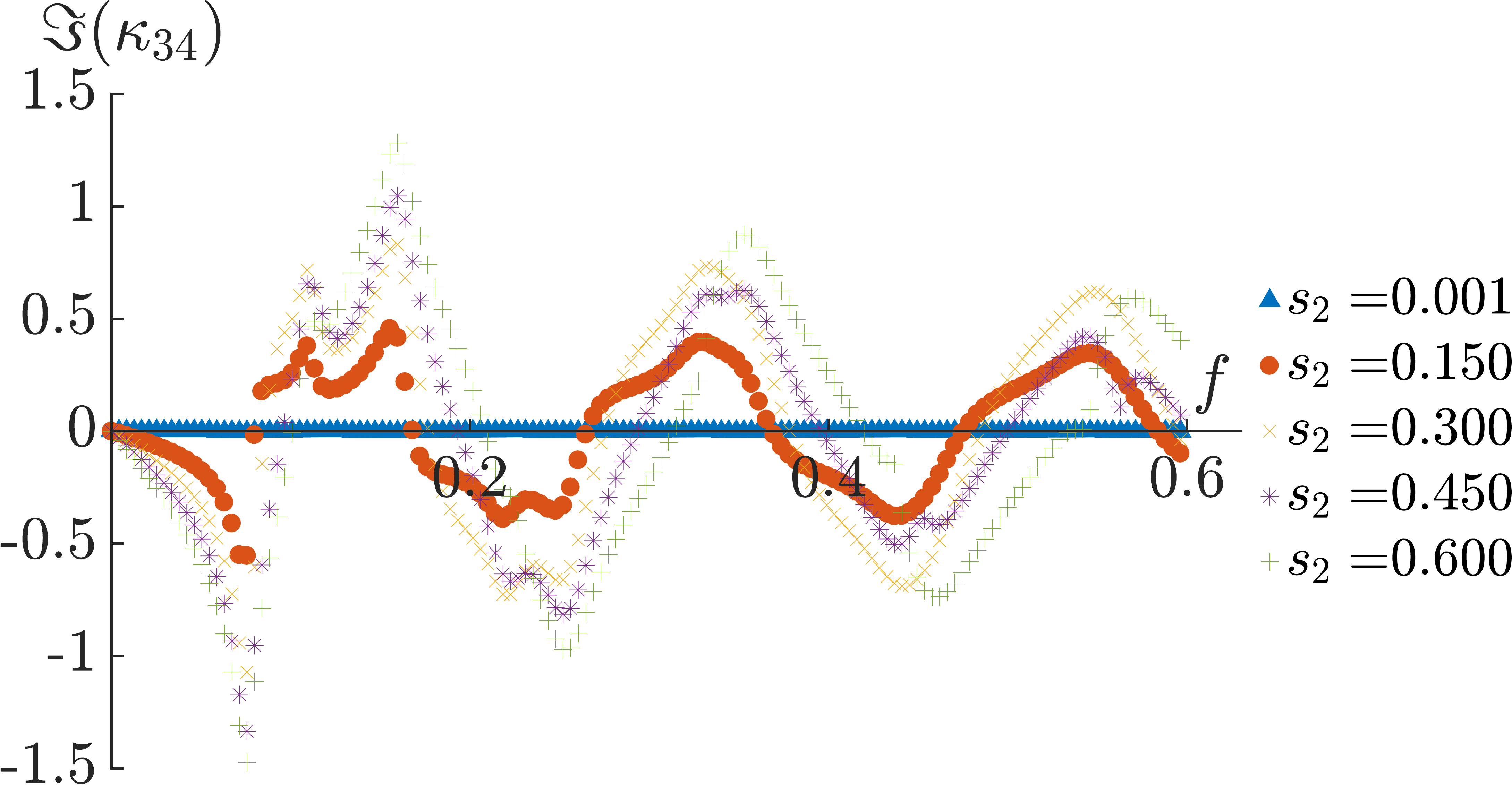}
		\caption{\label{fig:Im-ka34}}
	\end{subfigure}
	\caption{\label{fig:ka34}(\subref{fig:Re-ka34}) and (\subref{fig:Im-ka34}) show the real and imaginary parts of the coupling term, $\kappa_{34}$, as functions of frequency, $f$. Note that the negative of these values are also acceptable roots based on the present analysis.}
\end{figure}

\begin{figure}[!ht]
	\begin{subfigure}[b]{0.5\linewidth}
		\centering\includegraphics[height=100pt,left]{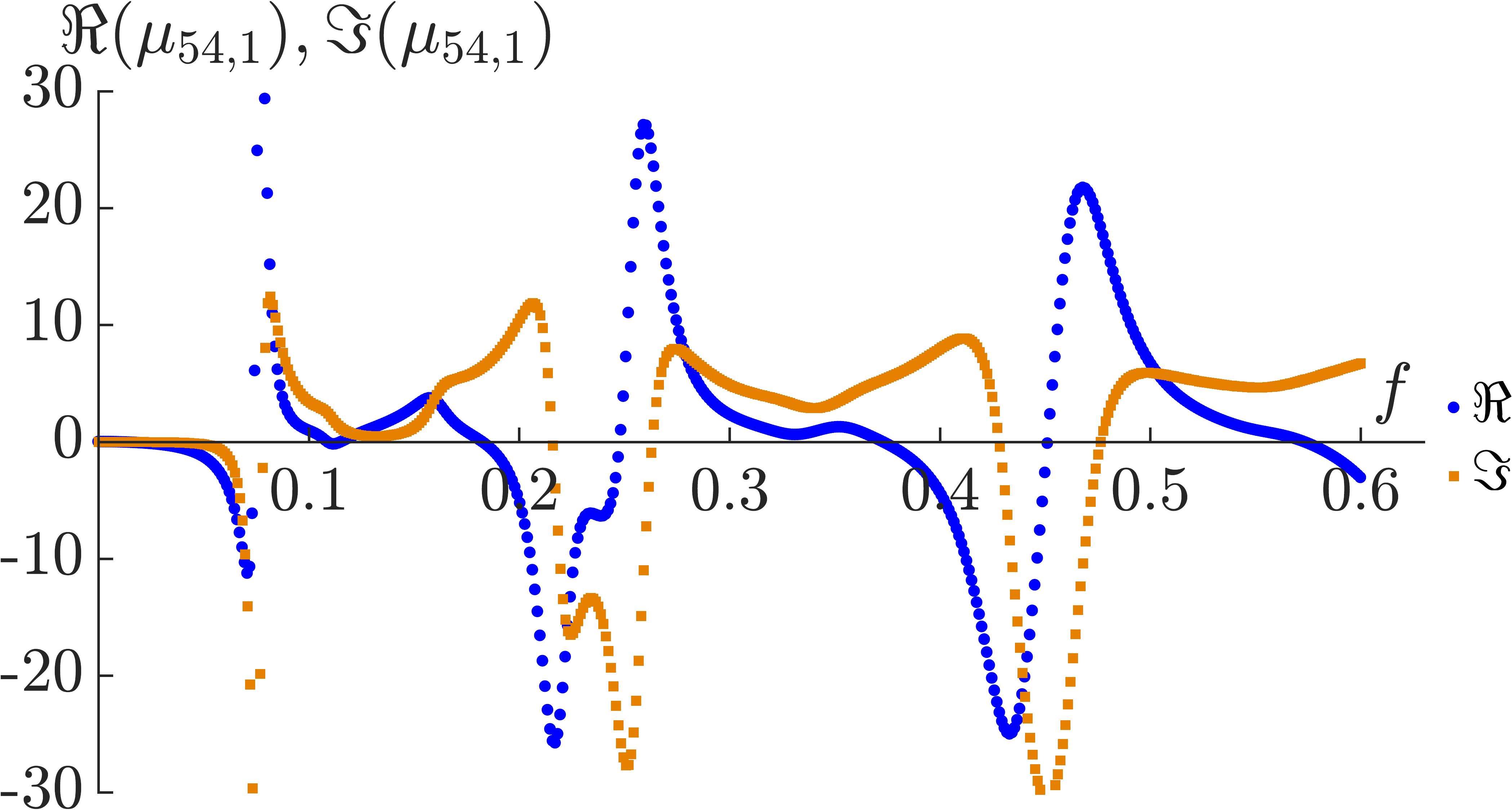}
		\caption{\label{fig:mu54os2}}
	\end{subfigure}
	\begin{subfigure}[b]{0.5\linewidth}
		\centering\includegraphics[height=100pt,left]{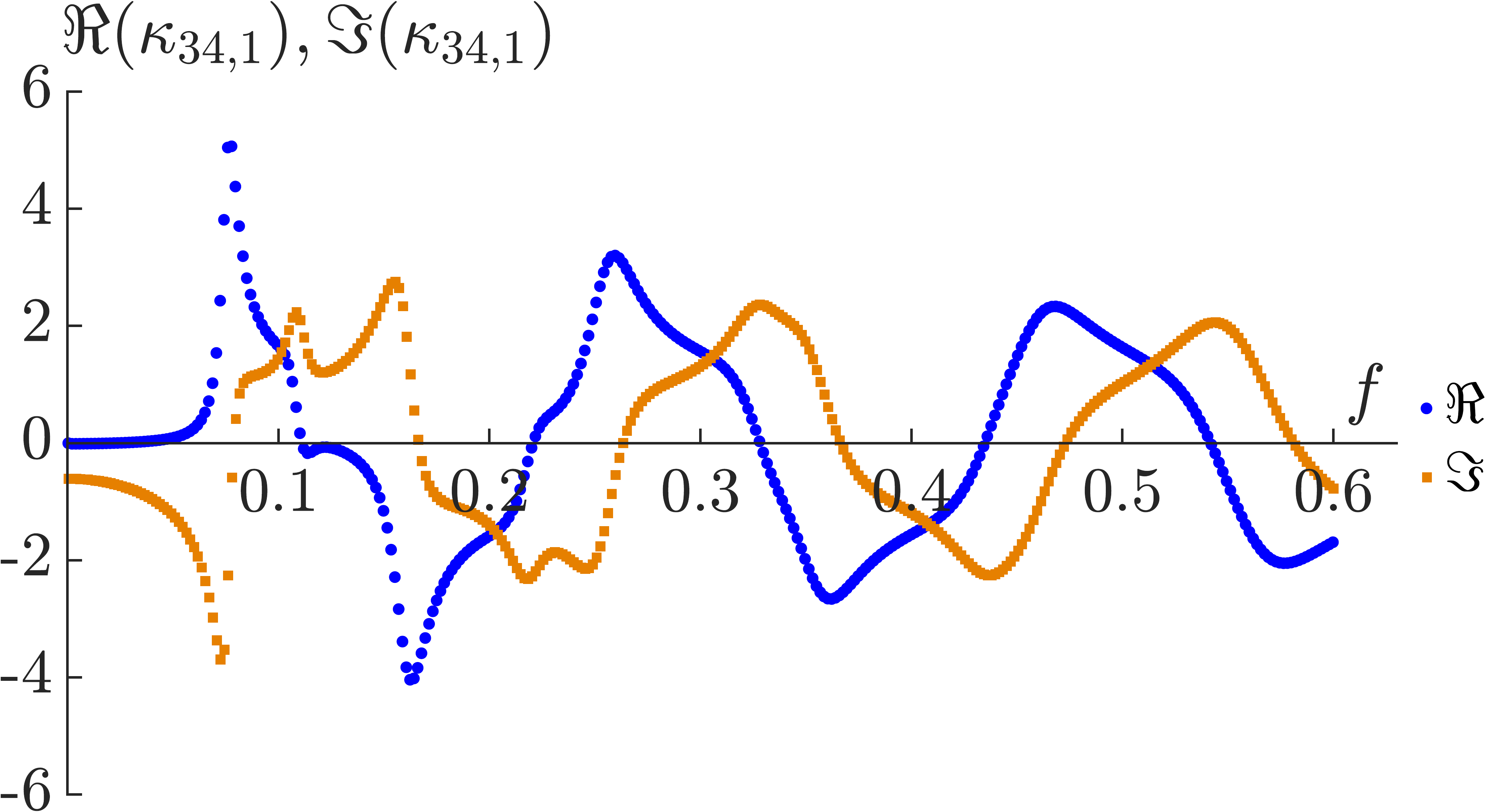}
		\caption{\label{fig:kap34os2}}
\end{subfigure}
	\caption{\label{fig:mu54os2kap34os2} The real and imaginary parts of the  (\subref{fig:mu54os2}) $\mu_{54,1}$ and  (\subref{fig:kap34os2}) $\kap_{34,1}$ (one root), as functions of frequency, $f$. Note the nonzero value of $\Im(\kap_{34,1})$ at zero frequency indicating the nonlocality of the system is not removable even as $f \rightarrow 0$ and $s_2 \rightarrow 0$.}
\end{figure}

For the case when $\mu_{44}$ and $\mu_{55}$ are assumed independent of the wave vector, their values match the $s_2 \rightarrow 0$ functions in the previous calculation. They also approach Voigt and Reuss averages of the shear moduli of the layers, respectively; See Figure~(\ref{fig:mu55an44}). However, in this scenario, $\mu_{54}$ is non-zero and a function of wave vector and can be calculated using  Equation ~(\ref{eq:mu54}). Figure~(\ref{fig:mu54}) illustrates the variations of $\mu_{54}$ with respect to frequency for different $s_2$ values. 
The coupling function $\kap_{34}$ is also non-zero, a function of the wave vector, and can be calculated from Equation~(\ref{eq:kap^2}). Note that there are two possible roots for $\kap_{34}$. For higher values of frequency, presence of branch cuts make it difficult to enforce continuity of $\kap_{34}$. The exact root selection may be resolved if $z_{43}$ is known.
There is, however, no ambiguity in $\kappa_{34}^2$ and the overall scattering and dispersion of the the medium is only dependent on it. One continuous root selection is shown in Figure~(\ref{fig:ka34}), while its negative is similarly acceptable in this analysis. $\mu_{54}$ and $\kap_{34}$ are odd function of $s_2$ and can get linearized for small enough $s_2$. The values of $\mu_{54,1}$ and $\kap_{34,1}$ (one root) from Equation~\eqref{eq:lin} are shown in Figure~(\ref{fig:mu54os2kap34os2}). Note that the nonzero value of $\Im(\kap_{34,1})$ at small frequencies indicate that the nonlocality of the system is inherently not removable even as $f \rightarrow 0$ and $s_2 \rightarrow 0$.

One can provide an estimate for $z_{43}$ based on the formulation here, even though the exact value is not accessible based on scattering measurements. In the diagonal tensor case, there is no ambiguity involved in calculating the $z_{43}$ based on $\mu_{44}$, while the exact value of $\kap_{34}$ (and not just its square) is needed in the other case. These estimates may be compared with the ones calculated using an integration scheme of the wave equation to further elucidate the acceptable mathematical approaches to overall properties determination of the physical heterogeneous problem.
\section{Summary and conclusions}
The scattering of oblique SH waves from a periodic layered slab may be calculated easily using the transfer matrix of such systems. It is shown here that one can use the scattering data to calculate overall constitutive tensors for any such symmetric slab within the coupled Willis formalism. However, the presence of micro-structure immediately necessitates the dependence of the constitutive tensors on the wave vector, even for simplest cases. Without further assumptions on the constitutive structure and/or other scattering analysis, one may not uniquely determine the full tensor. A number of physically and mathematically attractive assumptions are used here to demonstrate the process for a simple example. It is observed that not all potential selections of free parameters are analytically acceptable, e.g. $\kap_{35} = 0$ may be identically assumed for symmetric structures, while $\kap_{34} = 0$ leads to inconsistent results. Other 3D structural symmetries not considered here may further limit the potential constitutive descriptions. Interestingly, one can identify certain non-trivial points in the wave vector space, around which the system becomes nearly non-dispersive. Further analysis of such simple layered systems may require one or more of the following approaches. One may consider the scattering of coupled P/SV waves in a similar manner and utilize the expected symmetry of the response that demand more restrictive forms on the quantities that were discussed here. For example, the values of shear moduli for a transversely isotropic system are expected to match for both SH and P/SV waves. Alternatively, one can utilize field integration techniques to analyze through thickness quantities such as $\tau_4$ and $z_{43}$. Finally and in a sense related to the previous approach, one may consider the physical problem of scattering off surfaces other than the parallel plane faces of the layers. The last approach however brings an extra level of complexity as one has to consider non-specular scattering. 
\clearpage

\section*{REFERENCES}
\bibliographystyle{unsrt}
\bibliography{Manuscript}

\begin{thebibliography}{10}

\bibitem{Pitaevskii1984}
L.~D. Landau and E.~M. Lifshitz.
\newblock {\em {Electrodynamics of continuous media}}.
\newblock Pergamon Press, 2nd edition, 1984.

\bibitem{agranovich_crystal_1984}
V.~M. Agranovich and V.~Ginzburg.
\newblock {\em Crystal {Optics} with {Spatial} {Dispersion}, and {Excitons}}.
\newblock Springer {Series} in {Solid}-{State} {Sciences}. Springer-Verlag,
  Berlin Heidelberg, 2 edition, 1984.

\bibitem{Maradudin1973}
A.~A. Maradudin and D.~L. Mills.
\newblock {Effect of Spatial Dispersion on the Properties of a Semi-Infinite
  Dielectric}.
\newblock {\em Physical Review B}, 7(6):2787--2810, 1973.

\bibitem{Asadchy}
V.~S. Asadchy, M.~Albooyeh, S.~N. Tcvetkova, A.~D{\'{i}}az-Rubio, Y.~Ra'Di, and
  S.~A. Tretyakov.
\newblock {Perfect control of reflection and refraction using spatially
  dispersive metasurfaces}.
\newblock {\em Physical Review B}, 94(7), 2016.

\bibitem{Diaz-Rubio}
A.~D{\'{i}}az-Rubio, V.~S. Asadchy, A.~Elsakka, and S.~A. Tretyakov.
\newblock {From the generalized reflection law to the realization of perfect
  anomalous reflectors}.
\newblock {\em Science Advances}, 3(8), 2017.

\bibitem{Shalin2015}
A.~S. Shalin, P.~Ginzburg, A.~A. Orlov, I.~Iorsh, P.~A. Belov, Y.~S. Kivshar,
  and A.~V. Zayats.
\newblock {Scattering suppression from arbitrary objects in spatially
  dispersive layered metamaterials}.
\newblock {\em Physical Review B - Condensed Matter and Materials Physics},
  91(12), 2015.

\bibitem{Melnyk1968}
A.~R. Melnyk and M.~J. Harrison.
\newblock {Resonant Excitation of Plasmons in Thin Films by Elecromagnetic
  Waves}.
\newblock {\em Physical Review Letters}, 21(2):85--88, 1968.

\bibitem{JONES1969}
W.~E. Jones, K.~L. Kliewer, and R.~Fuchs.
\newblock {Nonlocal Theory of the Optical Properties of Thin Metallic Films}.
\newblock {\em Physical Review}, 178(3):1201--1203, 1969.

\bibitem{Melnyk1970}
A.~R. Melnyk and M.~J. Harrison.
\newblock {Theory of Optical Excitation of Plasmons in Metals}.
\newblock {\em Physical Review B}, 2(4):835--850, 1970.

\bibitem{Ruppin1975}
R.~Ruppin.
\newblock {Optical properties of small metal spheres}.
\newblock {\em Physical Review B}, 11(8):2871--2876, 1975.

\bibitem{Agranovich2006}
V.~M. Agranovich and Y.~N. Gartstein.
\newblock {Spatial dispersion and negative refraction of light}.
\newblock {\em Physics-Uspekhi}, 49(10), 2006.

\bibitem{Chern2013}
R.~Chern.
\newblock {Spatial dispersion and nonlocal effective permittivity for periodic
  layered metamaterials}.
\newblock {\em Optics Express}, 21(14):85--88, 2013.

\bibitem{AndreaAi2011}
A.~Al{\`{u}}.
\newblock {Restoring the physical meaning of metamaterial constitutive
  parameters}.
\newblock {\em Physical Review B - Condensed Matter and Materials Physics},
  83(8), 2011.

\bibitem{Alitalo2013}
P.~Alitalo, A.~E. Culhaoglu, C.~R. Simovski, and S.~A. Tretyakov.
\newblock {Experimental study of anti-resonant behavior of material parameters
  in periodic and aperiodic composite materials}.
\newblock {\em Journal of Applied Physics}, 113(22), 2013.

\bibitem{Li}
P.~Li, S.~Yao, X.~Zhou, G.~Huang, and G.~Hu.
\newblock {Effective medium theory of thin-plate acoustic metamaterials}.
\newblock {\em The Journal of the Acoustical Society of America},
  135(4):1844--1852, 2014.

\bibitem{Vehmas2014}
J.~Vehmas, S.~Hrabar, and S.~Tretyakov.
\newblock {Omega transmission lines with applications to effective medium
  models of metamaterials}.
\newblock {\em Journal of Applied Physics}, 115(13), 2014.

\bibitem{Papantonis}
S.~Papantonis, S.~Lucyszyn, and E.~Shamonina.
\newblock {Dispersion effects in Fakir's bed of nails metamaterial waveguides}.
\newblock {\em Journal of Applied Physics}, 115(5), 2014.

\bibitem{Belov2003}
P.~A. Belov, R.~Marqu{\'{e}}s, S.~I. Maslovski, I.~S. Nefedov, M.~Silveirinha,
  C.~R. Simovski, and S.~A. Tretyakov.
\newblock {Strong spatial dispersion in wire media in the very large wavelength
  limit}.
\newblock {\em Physical Review B}, 67(11), 2003.

\bibitem{Rockstuhl2008}
C.~Rockstuhl, C.~Menzel, T.~Paul, T.~Pertsch, and F.~Lederer.
\newblock {Light propagation in a fishnet metamaterial}.
\newblock {\em Physical Review B}, 78(15), 2008.

\bibitem{Tyshetskiy2014}
Y.~Tyshetskiy, S.~V. Vladimirov, A.~E. Ageyskiy, I.~I. Iorsh, A.~Orlov, and
  P.~A. Belov.
\newblock {Guided modes in a spatially dispersive wire medium slab}.
\newblock {\em Journal of the Optical Society of America B}, 31(8), 2014.

\bibitem{Miret2016}
J.~J. Miret, J.~{Aitor Sorn{\'{i}}}, M.~Naserpour, A.~{Ghasempour Ardakani},
  and C.~J. Zapata-Rodr{\'{i}}guez.
\newblock {Nonlocal dispersion anomalies of Dyakonov-like surface waves at
  hyperbolic media interfaces}.
\newblock {\em Photonics and Nanostructures - Fundamentals and Applications},
  18:16--22, 2016.

\bibitem{Yves}
S.~Yves, R.~Fleury, T.~Berthelot, M.~Fink, F.~Lemoult, and G.~Lerosey.
\newblock {Crystalline metamaterials for topological properties at
  subwavelength scales}.
\newblock {\em Nature Communications}, 8, 2017.

\bibitem{Ushkov}
A.~A. Ushkov and A.~A. Shcherbakov.
\newblock {Concurrency of anisotropy and spatial dispersion in low refractive
  index dielectric composites}.
\newblock {\em Optics Express}, 25(1), 2017.

\bibitem{Hopfield1963}
J.~J. Hopfield and D.~G. Thomas.
\newblock {Theoretical and experimental effects of spatial dispersion on the
  optical properties of crystals}.
\newblock {\em Physical Review}, 132(2):563--572, 1963.

\bibitem{Portigal}
D.~L. Portigal and E.~Burstein.
\newblock {Acoustical activity and other first-order spatial dispersion effects
  in crystals}.
\newblock {\em Physical Review}, 170(3):673--678, 1968.

\bibitem{Tretyakov1998}
S.~A. Tretyakov.
\newblock {Uniaxial Omega Medium as a Physically Realizable Alternative for the
  Perfectly Matched Layer (Pml)}.
\newblock {\em Journal of Electromagnetic Waves and Applications},
  12(6):821--837, 1998.

\bibitem{Ciattoni2015}
A.~Ciattoni and C.~Rizza.
\newblock {Nonlocal homogenization theory in metamaterials: Effective
  electromagnetic spatial dispersion and artificial chirality}.
\newblock {\em Physical Review B - Condensed Matter and Materials Physics},
  91(18), 2015.

\bibitem{Agarwal}
G.~S. Agarwal, D.~N. Pattanayak, and E.~Wolf.
\newblock {Electromagnetic fields in spatially dispersive media}.
\newblock {\em Physical Review B}, 10(4):1447--1475, 1974.

\bibitem{Puri1983}
A.~Puri and J.~L. Birman.
\newblock {Pulse propagation in spatially dispersive media}.
\newblock {\em Physical Review A}, 27(2):1044--1052, 1983.

\bibitem{Belov}
P.~A. Belov, C.~R. Simovski, and S.~A. Tretyakov.
\newblock {Example of bianisotropic electromagnetic crystals: The spiral
  medium}.
\newblock {\em Physical Review E}, 67(5), 2003.

\bibitem{Yaghjian2013}
A.~D. Yaghjian, A.~Al{\`{u}}, and M.~G. Silveirinha.
\newblock {Homogenization of spatially dispersive metamaterial arrays in terms
  of generalized electric and magnetic polarizations}.
\newblock {\em Photonics and Nanostructures - Fundamentals and Applications},
  11(4):374--396, 2013.

\bibitem{Popa2009}
B.~Popa and S.~A. Cummer.
\newblock {Design and characterization of broadband acoustic composite
  metamaterials}.
\newblock {\em Physical Review B}, 80(17), 2009.

\bibitem{Castanie2014}
A.~Castani{\'{e}}, J.~F. Mercier, S.~F{\'{e}}lix, and A.~Maurel.
\newblock {Generalized method for retrieving effective parameters of
  anisotropic metamaterials}.
\newblock {\em Optics Express}, 22(24):29937--29953, 2014.

\bibitem{Park2016}
J.~H. Park, H.~J. Lee, and Y.~Y. Kim.
\newblock {Characterization of anisotropic acoustic metamaterial slabs}.
\newblock {\em Journal of Applied Physics}, 119(3), 2016.

\bibitem{Lafarge2013}
D.~Lafarge and N.~Nemati.
\newblock {Nonlocal Maxwellian theory of sound propagation in fluid-saturated
  rigid-framed porous media}.
\newblock {\em Wave Motion}, 50(6):1016--1035, 2013.

\bibitem{Nemati2014}
N.~Nemati and D.~Lafarge.
\newblock {Check on a nonlocal Maxwellian theory of sound propagation in
  fluid-saturated rigid-framed porous media}.
\newblock {\em Wave Motion}, 51(5):716--728, 2014.

\bibitem{Nemati2015}
N.~Nemati, A.~Kumar, D.~Lafarge, and N.~X. Fang.
\newblock {Nonlocal description of sound propagation through an array of
  Helmholtz resonators}.
\newblock {\em Comptes Rendus M{\'{e}}canique}, 343(343):656--669, 2015.

\bibitem{nemati_nonlocal_2017}
Navid Nemati, Yoonkyung~E. Lee, Denis Lafarge, Aroune Duclos, and Nicholas
  Fang.
\newblock Nonlocal dynamics of dissipative phononic fluids.
\newblock {\em Physical Review B}, 95(22), June 2017.

\bibitem{Lee2016}
H.~J. Lee, H.~S. Lee, P.~S. Ma, and Y.~Y. Kim.
\newblock {Effective material parameter retrieval of anisotropic elastic
  metamaterials with inherent nonlocality}.
\newblock {\em Journal of Applied Physics}, 120(10), 2016.

\bibitem{Amirkhizi2017}
A.~V. Amirkhizi.
\newblock {Homogenization of layered media based on scattering response and
  field integration}.
\newblock {\em Mechanics of Materials}, 114:76--87, 2017.

\bibitem{Nemat-Nasser2015a}
S.~Nemat-Nasser.
\newblock {Anti-plane shear waves in periodic elastic composites: band
  structure and anomalous wave refraction}.
\newblock {\em Proceedings of the Royal Society A: Mathematical, Physical and
  Engineering Science}, 471(2180), 2015.

\bibitem{Srivastava2016}
A.~Srivastava.
\newblock {Metamaterial properties of periodic laminates}.
\newblock {\em Journal of the Mechanics and Physics of Solids}, 96:252--263,
  2016.

\bibitem{Vinh2015}
P.~C. Vinh, T.~T. Tuan, and M.~A. Capistran.
\newblock {Explicit formulas for the reflection and transmission coefficients
  of one-component waves through a stack of an arbitrary number of layers}.
\newblock {\em Wave Motion}, 54:134--144, 2015.

\bibitem{Milton2007}
G.~W. Milton and J.~R. Willis.
\newblock {On modifications of Newton's second law and linear continuum
  elastodynamics}.
\newblock {\em Proceedings of the Royal Society A: Mathematical, Physical and
  Engineering Sciences}, 463(2079):855--880, 2007.

\bibitem{Willis2009}
J.~R. Willis.
\newblock {Exact effective relations for dynamics of a laminated body}.
\newblock {\em Mechanics of Materials}, 41(4):385--393, 2009.

\bibitem{Nantasetphong2018}
W.~Nantasetphong, Z.~Jia, M.A. Hasan, A.V. Amirkhizi, and S.~Nemat-Nasser.
\newblock A new technique for characterization of low impedance materials at
  acoustic frequencies.
\newblock {\em Experimental Mechanics}, Jul 2018.

\end{thebibliography}

\end{document}